\documentstyle[emulateapj,psfig]{article}

\received{}
\revised{}
\accepted{}

\cpright{}{}
\journalid{}{}
\articleid{}{}
\paperid{}

\cpright{}{}
\ccc{}

\lefthead{Alcock et al. (The {\sc macho} collaboration)}
\righthead{FREQUENCY ANALYSIS OF THE RR LYRAE STARS IN THE LMC}

\begin{document}

\title{THE MACHO PROJECT LARGE MAGELLANIC CLOUD VARIABLE STAR INVENTORY. 
XI. FREQUENCY ANALYSIS OF THE FUNDAMENTAL MODE RR LYRAE STARS
}

\author{
Alcock,~C.$^{2,3,4}$,  
Alves,~D. R.$^{6}$, 
Becker,~A.$^{8}$, 
Bennett,~D.$^{3,9}$, 
Cook,~K.H.$^{3,4}$, 
Drake,~A.,$^{3,10}$, 
Freeman,~K.$^{11}$, 
Geha,~M.$^{1}$, 
Griest,~K.$^{12}$, 
Kov\'acs,~G.$^{5}$, 
Lehner,~M.$^{2}$, 
Marshall,~S.$^{3}$, 
Minniti,~D.$^{10}$, 
Nelson,~C.$^{3,13}$, 
Peterson,~B.$^{11}$, 
Popowski,~P.$^{14}$, 
Pratt,~M.$^{15}$, 
Quinn,~P.$^{16}$, 
Rodgers,~A.$^{7}$, 
Stubbs,~C.$^{4,15}$, 
Sutherland,~W.$^{17}$,  
Vandehei,~T.$^{12}$ 
and Welch,~D.L.$^{18}$}

\footnotetext[1]{Department of Astronomy and Astrophysics,
    University of California, Santa Cruz 95064; 
    mgeha@ucolick.org}
\footnotetext[2]{Department of Physics and Astronomy, University of
    Pennsylvania, Philadelphia, PA, 19104-6396; 
    alcock, mlehner@hep.upenn.edu}
\footnotetext[3]{Lawrence Livermore National Laboratory, Livermore, CA 94550; 
    kcook, adrake, stuart@igpp.ucllnl.org, nelson61@llnl.gov}
\footnotetext[4]{Center for Particle Astrophysics, University of California,
    Berkeley, CA 94720}
\footnotetext[5]{Konkoly Observatory, P.O. Box 67, H-1525 Budapest, Hungary; 
    kovacs@konkoly.hu}
\footnotetext[6]{Columbia Astrophysics Laboratory,
    MailCode 5247, 550 W. 120th St., New York, NY 10027; 
    alves@astro.columbia.edu}
\footnotetext[7]{deceased}
\footnotetext[8]{Bell Laboratories, Lucent Technologies, 600 Mountain Avenue, 
    Murray Hill, NJ 07974; 
    acbecker@physics.bell-labs.com}
\footnotetext[9]{Department of Physics, University of Notre Dame, IN 46556; 
    bennett@bustard.phys.nd.edu}
\footnotetext[10]{Depto. de Astronomia, P. Universidad Catolica, Casilla 306,
    Santiago 22, Chile; 
    dante@astro.puc.cl}
\footnotetext[11]{Res. School of Astron. and Astrphys., Mount Stromlo Obs., 
    Cotter Road, Weston, ACT 2611, Australia; kcf, peterson@mso.anu.edu.au}
\footnotetext[12]{Department of Physics, University of California,
    San Diego, CA 92093; 
    kgriest@ucsd.edu, vandehei@astrophys.ucsd.edu}
\footnotetext[13]{Department of Physics, University of California, Berkeley,
    CA 94720}
\footnotetext[14]{MPA, Karl-Schwarzschild-Str. 1, Pf. 1317, 85741 Garching 
    b. M\"unchen, Germany; popowski@mpa-garching.mpg.de}
\footnotetext[15]{Departments of Astronomy and Physics,
    University of Washington, Seattle, WA 98195; 
    stubbs@astro.washington.edu}
\footnotetext[16]{European Southern Observatory, Karl Schwarzchild Str.\ 2,
    D-8574 8 G\"{a}rching bel M\"{u}nchen, Germany;    
    pjq@eso.org}
\footnotetext[17]{Department of Physics, University of Oxford,
    Oxford OX1 3RH, U.K.; 
    w.sutherland@physics.ox.ac.uk}
\footnotetext[18]{Department of Physics and Astronomy, McMaster University,
    Hamilton, Ontario, Canada, L8S 4M1; 
    welch@physics.mcmaster.ca}

\begin{abstract}
We have frequency analyzed 6391 variables classified earlier as 
fundamental mode RR~Lyrae (RR0) stars in the {\sc macho} database 
on the Large Magellanic Cloud (LMC). The overwhelming majority 
(i.e., 96\%) of these variables have been proved to be indeed RR0 
stars, whereas the remaining ones have fallen in one of the 
following categories: single- and double-mode Cepheids, binaries, 
first overtone and double-mode RR~Lyrae stars and non-classified 
variables. Special attention has been paid to the properties of the 
amplitude- and phase-modulated RR0 stars (the Blazhko stars). We 
found altogether 731 Blazhko variables showing either a doublet or 
an equidistant triplet pattern at the main pulsation component in 
their frequency spectra. This sample overwhelmingly exceeds the 
number of Blazhko stars known in all other systems combined. The 
incidence rate of the Blazhko variables among the fundamental mode 
RR~Lyrae stars in the LMC is 11.9\%, which is three times higher 
than their rate among the first overtone RR~Lyrae stars. No difference 
is found in the average brightness between the single-mode and 
Blazhko variables. However, the latter ones show a somewhat lower 
degree of skewness in their average light curves and a concomitant 
lower total amplitude in their modulation-free light curves. From 
the frequency spectra we found that variables with larger modulation 
amplitudes at the higher frequency side of the main pulsation component 
are three times more frequent than the ones showing the opposite 
amplitude pattern. A search made for a modulation component with 
the Blazhko period in the average brightness of the individual 
variables showed the existence of such a modulation with an overall 
amplitude of $\approx 0.006$~mag. On the other hand, a similar 
search for quadruple modulation patterns around the main pulsation 
component have failed to clearly detect such components at the 
$\approx 0.004$~mag level. This means that the amplitudes of the 
quadruple components (if they exist) should be, on the average, at 
least ten times smaller than those of the triplet components. This 
finding and the existence of Blazhko variables with highly asymmetric 
modulation amplitudes not only question the validity of the magnetic 
oblique rotator model, but also puts stringent constraints on models 
based on mode coupling theories.      
\end{abstract} 
\keywords{
globular clusters: general ---
stars: horizontal-branch ---
stars: oscillations --- 
stars: variables: other (RR~Lyrae)
}

%
%

\section{INTRODUCTION}

The mystery of the physical cause of amplitude-modulation
in certain RR Lyrae stars, popularly known as the ``Blazhko Effect''
remains despite almost a century of study. In a previous paper in 
this series (Alcock et al. 2000, hereafter A00) we discussed the 
results of the frequency analysis of a sample of 1350 Large 
Magellanic Cloud (LMC) variables, formerly classified as first 
overtone RR~Lyrae (RR1) stars. In addition to the discovery of more 
than one hundred double-mode variables, we found 52 stars which 
displayed long-period amplitude and phase modulations. Until very 
recently (Olech, Ka{\l}u\.{z}ny \& Thompson 2001, and references 
therein; Cseresnjes 2001; Moskalik \& Poretti 2003; Soszy\'{n}ski 
et al. 2003) it was established as a well-observed phenomenon only 
among fundamental-mode variables. The study of A00 gave the first 
reliable statistics on the incidence rate of this behavior among 
RR1 stars. A rate of 4\% was found which is surprisingly low compared 
with the most often referenced rate of 20--30\% for the fundamental 
mode (RR0) stars (Szeidl 1988). The main purpose of the present study 
is to verify whether this large difference does indeed exist between 
these two groups of stars. The large and homogeneous sample of the 
{\sc macho} database stars used in this study yield much more 
significant results for the global statistical properties of this 
population than any other existing investigation which were typically 
based on very limited and inhomogeneous samples of objects from various 
stellar systems. (Exceptions to this characterization are the surveys 
of Soszy\'{n}ski et al. 2003 on the 7612 RR~Lyrae stars of the LMC from 
the {\sc ogle-ii} database, of Moskalik \& Poretti 2003 for the 215 
variables of the Galactic Bulge from the {\sc ogle-i} database and of 
Cseresnjes 2001 for the 3700 variables of the Fornax Dwarf galaxy.)
  
Apart from the task of identification of the variable type for 
RR~Lyrae in the LMC {\sc macho} database, we also study the global 
properties of the Blazhko stars and compare them with their 
singly-periodic counterparts. A key discriminant between rival 
explanations for the Blazhko behavior is found in the predicted 
frequency and amplitude pattern distributions of the amplitude 
spectra of these stars. The existence (or non-existence) of 
triplet or quintuplet structures and the degree of asymmetry of 
the corresponding modulation amplitudes strongly restricts the 
range of acceptable models. The present large dataset enables us 
to put significant constraints on the possible ranges of the various 
modulation components. A brief summary of the results obtained by 
this analysis of a fraction of the present dataset has already been 
published by Welch et al. (2002).   
    
%
%

\section{THE DATA, THE METHOD OF ANALYSIS AND VARIABLE CLASSIFICATION}

The dataset analyzed in this paper comprises 6391 light curves 
selected from the {\sc macho} database as variables satisfying 
certain conditions on brightness, color and period, corresponding 
to the expected properties of fundamental mode RR~Lyrae stars in 
the LMC\footnote {
The selection of the RR Lyrae for this
paper was made many years before the completion of the MACHO
Project. In the time since the submission of this paper, we
have become aware that there are no RR Lyrae variables in the
eastern half of field 5 and an underabundance in a small number
of other locations. This was apparently due to an access problem
of the program used to phase the photometry. Our ability to test
this hypothesis end-to-end has been hampered by the unavailability
of scripts present on Mount Stromlo Observatory computers which
were lost during the devastating fire of Junuary 18, 2003.
We have since extracted a more complete list of RR Lyrae from the
full eight-year database and we will make this list available to
interested investigators. It is our strong expectation that the
frequency of Blazhko behavior reported in this paper will not be
significantly altered by a similar analysis of the additional
candidates in the eight-year list.
}. 
Due to overlapping fields, 364 light curves turned out to 
represent variables already appearing in the database with other 
identification numbers. Since these double (or multiple) 
identifications decrease the number of stars in each variable 
type by only a few percent, and therefore, does not influence 
the incidence rates in a significant way, all of our subsequent 
results and statistics refer to the full sample of 6391 
identifications. The list of stars with multiple identifications 
is presented at the end of this section.

This paper analyzes RR Lyrae stars identified in 30 LMC {\sc macho} 
Project fields (1, 2, 3, 5, 6, 7, 9, 10, 11, 12, 13, 14, 15, 17, 18, 
19, 22, 23, 24, 47, 53, 55, 57, 76, 77, 78, 79, 80, 81, 82). These 
fields include 54\% of the observed stars in all fields because they 
cover the densely populated bar of the LMC 
(see the home page\footnote {http://www.macho.mcmaster.ca} of the 
{\sc macho} Project). The data 
utilized span a period of 2700~d (7.4~yr), which is 1~yr longer 
than the timebase of the first overtone RR~Lyrae stars analyzed in 
A00. The magnitudes have been transformed to the Johnson V system 
according to the method described by Alcock et al. (1999). As a result 
of the longer time baseline, the average number of observation epochs 
per star is also greater, often in the range 700 to 1000. The sampling 
rate and data distribution is the same as in the case of the RR1 stars. 
In particular, there is only a 1~d quasi-periodicity in the sampling 
but no trace of 1~yr aliasing. The accuracy of the individual data 
points is also the same as for the RR1 data, but, because of the 
larger amplitudes, the overall signal-to-noise ratio (SNR=$A/\sigma$, 
$A$ is the total (peak-to-peak) amplitude, $\sigma$ is the standard 
deviation of the residuals) is larger ($\langle$SNR$\rangle \approx 11$, 
whereas for the RR1 stars $\langle$SNR$\rangle \approx 2.6$, although 
this figure refers to the data in the 'r' color, which further lowers 
the SNR -- see A00).

%
%
\begin{deluxetable}{llrrc}
\tablewidth{0pc}
\tablecaption{Variable types in the {\sc macho} LMC database}
\tablehead{
{\it Type} & {\it Short description} & {\it Number} & \% 
           & $\sigma$(\%) }
\startdata
RR0$-$S     & Singly-periodic FU RR~Lyrae              & 4882 & 79.3 & 0.5 \\
RR0$-$BL1   & RR0 with 1 close frequency component     &  400 &  6.5 & 0.3 \\
RR0$-$BL2   & RR0 with 2 close symmetric frequency components  
                                                       &  331 &  5.4 & 0.3 \\ RR0$-\nu$M  & RR0 with several close components        &   20 &  0.3 & 0.1 \\
RR0$-$PC    & RR0 with period change                   &  177 &  2.9 & 0.2 \\
RR0$-$D     & RR0 \& integer d$^{-1}$ frequencies      &  308 &  5.0 & 0.3 \\
RR0$-$MI    & RR0 with some miscellany                 &   40 &  0.6 & 0.1 \\
RR1         & Singly-periodic FO RR~Lyrae              &   36 &  $-$ & $-$ \\
RR01        & FU/FO RR~Lyrae                           &    6 &  $-$ & $-$ \\
NC          & Non-classified variables                 &   32 &  $-$ & $-$ \\
SOCEP       & Singly-periodic overtone Cepheid         &   60 &  $-$ & $-$ \\
FU/FO       & FU/FO double-mode Cepheid                &    2 &  $-$ & $-$ \\
FO/SO       & FO/SO double-mode Cepheid                &   58 &  $-$ & $-$ \\
BI          & Eclipsing binary                         &   39 &  $-$ & $-$ \\                     \enddata
\tablecomments{The symbol $\sigma$(\%) denotes the standard 
deviation of the population ratio, assuming Poisson distribution 
(see Equation (1)). In computing population ratios, only fundamental 
mode RR~Lyrae stars have been considered.} 
\end{deluxetable}

Apart from spectrum averaging (see Sect.~5), the method of analysis 
in this survey is essentially the same as that in A00. We performed 
a standard frequency analysis based on a Fourier approach and a
`pre-whitening' technique (e.g., Deeming 1975, with the fast 
transformation method recommended by Kurtz 1985). Before executing 
the Fourier transformation, outlying points have been omitted by 
checking the distribution of the magnitudes and leaving out data 
points separated from a compact set which is thought to represent 
the lightcurve plus random noise. The relative number of points 
omitted in this way varied between 1\% and 4\% and most often 
these omitted points were obtained in very poor seeing. The 
classification of the variables was performed in three major steps:
\begin{itemize}
\item
Frequency analysis with pre-whitening in the $[1,6]$~d$^{-1}$ band, 
visual inspection of the frequency spectra and folded lightcurves, 
and preliminary selection of variable types.
\item
Reanalysis of the pre-selected multiperiodic variables by three 
successive pre-whitenings and a closer examination of the frequency 
spectra.
\item
Visual inspection of the lightcurves of all variables to check 
the above classifications.
\end{itemize} 
Except for the first pre-whitening cycle, when the main pulsation 
component is filtered out from the time series, we used a 
single-component least squares fit for whitening. Because of the 
high nonlinearity, for the main pulsation component we utilized 
all harmonics up to the third one.
 
It is important to emphasize that the classification is based on the 
observed frequency pattern in a wide frequency band. Therefore, the 
significance of a given pattern is determined through a comparison of  
a large number of other features which might make that pattern less 
significant than if we made the comparison in a smaller frequency band 
(the chance of detecting higher, noise-induced peaks increases with 
the increase of the frequency band). For example, if we choose a narrow 
band around the main pulsation component, and search for equidistant 
triplet in the spectrum, a considerable fraction of the Blazhko (BL) 
stars that were classified after the inspection of the wide frequency 
band as BL stars with doublet patterns (hereafter BL1 stars) will prove 
to have triplet patterns and will be classified as BL2 stars 
(see Sect. 4). Although some arguments can be given for the preference 
of a classification based on pattern search in a narrow frequency band, 
we follow a more conservative approach, and our classification (if not 
stated otherwise) refer to the results obtained from the above wide 
frequency band. 

Variable types and their respective totals obtained in the course of 
the above analysis are listed in Table 1. The notation is the same as 
in A00, except for the Blazhko stars, where we use similar symbols both 
for variables with apparently single modulation components (BL1 stars) 
and those with two symmetrically-spaced modulation components straddling
the main pulsation component (BL2 stars). In A00 we assigned the symbol 
``BL'' only to the BL2 stars, in order to maintain a coherent nomenclature 
with the handful of variables which have been analyzed in the Galactic 
field. These few variables all display triplets in their frequency 
spectra. However the analysis of our larger sample strongly suggests 
that the two types of modulation overlap and they are indicative of the 
same underlying phenomenon which we denote by ``BL''.

As discussed earlier, our statistics have not been corrected for
multiple identifications due to field overlap. This results
in incidence rates which differ from the true rates by, at most, 
0.1\%. One reason that we did not make this correction is that, 
in critical cases, classification of the same variable star may 
depend on the particular time series we analyzed. For example, 
the same star might appear in one field with many observed epochs 
and in the other with a much smaller number. If a modulation component 
has an amplitude near the noise level, it could potentially be 
classified as both a Blazhko and as a single-mode variable (an example 
is with variable star identified as both, {\sc macho} ID 17.3194.3230 
and 10.3194.499). A similar ambiguity might arise in the classification 
of BL1 and BL2 stars if the signal-to-noise ratio is low (e.g., 
{\sc macho} ID 6.6811.969 and 13.6811.4172). We found five variables 
which have been classified both as BL1- and BL2-types. Among the 400 
BL1 stars there are 13 duplicates with the same BL1 classification. 
Similarly, we found 8 duplicate identifications among the 331 BL2 
stars which were classified identically in both instances. These 
duplications suggest that our classification is repeatable and 
reliable.

The notation in Table~1 incorporates the notation of Udalski et al. 
(1999) for the radial pulsation mode, in which the symbols FU, FO 
and SO denote the fundamental, first and second overtone modes, 
respectively. The RR0$-\nu$M variables are similar to the BL stars 
in that their amplitude spectra exhibit close secondary components. 
However, their frequency patterns are more complicated and show no 
obvious sign of dominant, equidistant frequency components (around 
the main pulsation component). For PC variables, the structure remaining 
around the main pulsation component cannot be resolved, and the few 
successive pre-whitenings do not eliminate these features from the 
amplitude spectrum. Although the failure of the pre-whitening 
technique indicates a long-term variation in the signal frequency 
(or even in the amplitude), we must note that it is also possible 
(at least in some cases) that we are witnessing Blazhko modulations 
with very long periods. This contention is supported by the fact 
that the present sample contains BL stars with modulation periods 
of several years. With this long BL periods, the modulation components 
are quite close to each other but are still well-resolved due to the 
long temporal baseline of the observations. Several BL stars 
additionally display a PC character. In such cases the variables 
have been classified as BL-type, despite the additional PC behavior.

A significant fraction (5\%) of the stars in this sample display
frequency components at integer $d^{-1}$. These are labelled RR0$-$D 
variables (such stars also were found in A00). Clearly this behavior 
indicates some instrumental or reduction artefact. At present the 
source of this is unknown.

The difference between the RR0$-$MI and NC variables is that in the 
former class (most probably) all variables are RR0 stars, but they 
exhibit extra peaks in their frequency spectra, which are not easy 
to explain within any simple framework (e.g., multimode pulsation). 
In the case of NC stars, classification is not possible either 
because of the limited dataset, or because of the `strange-looking' 
light curve (e.g., drift-like mean-light variation). While additional
interesting physical behavior may be present in these stars, we expect
(but have not established) that extrinsic effects are responsible for 
many such classifications.
   
Many new LMC Cepheid variables have already been discovered, both by 
the {\sc macho} and by the {\sc ogle} Projects. After 
cross-identification of the apparent Cepheid variables found in our 
initial sample with both databases, we find that 67\% of the FO/SO 
Cepheids reported in this paper are new discoveries. 

Statistical errors (standard deviations $\sigma$(\%)) of the 
incidence rates are calculated from the assumption that the 
population follows a Poisson distribution (i.e., $\sigma_i = \sqrt{N_i}$, 
where $N_i$ is the number of objects in the $i$-th population). 
This yields the following expression for the standard deviation of 
the population ratio $N_i/N$
%
%
%
\begin{eqnarray}
\sigma_{N_i/N} = {1\over \sqrt{N}}
\sqrt{{N_i\over N}\Big(1-{N_i\over N}\Big)} \hskip 2mm .
\end{eqnarray}
BL stars constitute a significant fraction of the RR0 population in 
the LMC according to Table~1. Therefore, for the first time, BL stars 
can be used for reliable statistical studies. These results will be 
presented in the next sections.

Since the spatial distribution of variables is a separate issue from 
the ones to be discussed in the following sections, here, in Figure~1, 
we show the field-by-field distribution of the BL and all RR0 stars. 
It is seen that there are considerable differences in the number of 
RR0 stars among the various fields, but this correlates with the 
stellar density of the fields (e.g., fields 53, 55 and 57 are far 
from the LMC bar). The ratio of BL to total number of RR0 stars also 
shows some fluctuations, especially in the lower-density fields. 
Therefore, this fluctuation might have a statistical origin. For 
example, for a sample size of N=50, an observed incidence rate of 
3\% has a 2.4\% standard deviation. In the well-populated fields 
the relative number ratios fluctuate around 10\%, suggesting that 
this ratio might be the same in all fields. When the full sample is 
considered, we get an average incidence rate of BL stars of 
$(11.9\pm 0.4)$\% (the error corresponds to $1\sigma$ standard 
deviation of the mean). 
%
%
%
\vskip 3mm
\centerline{\psfig{file=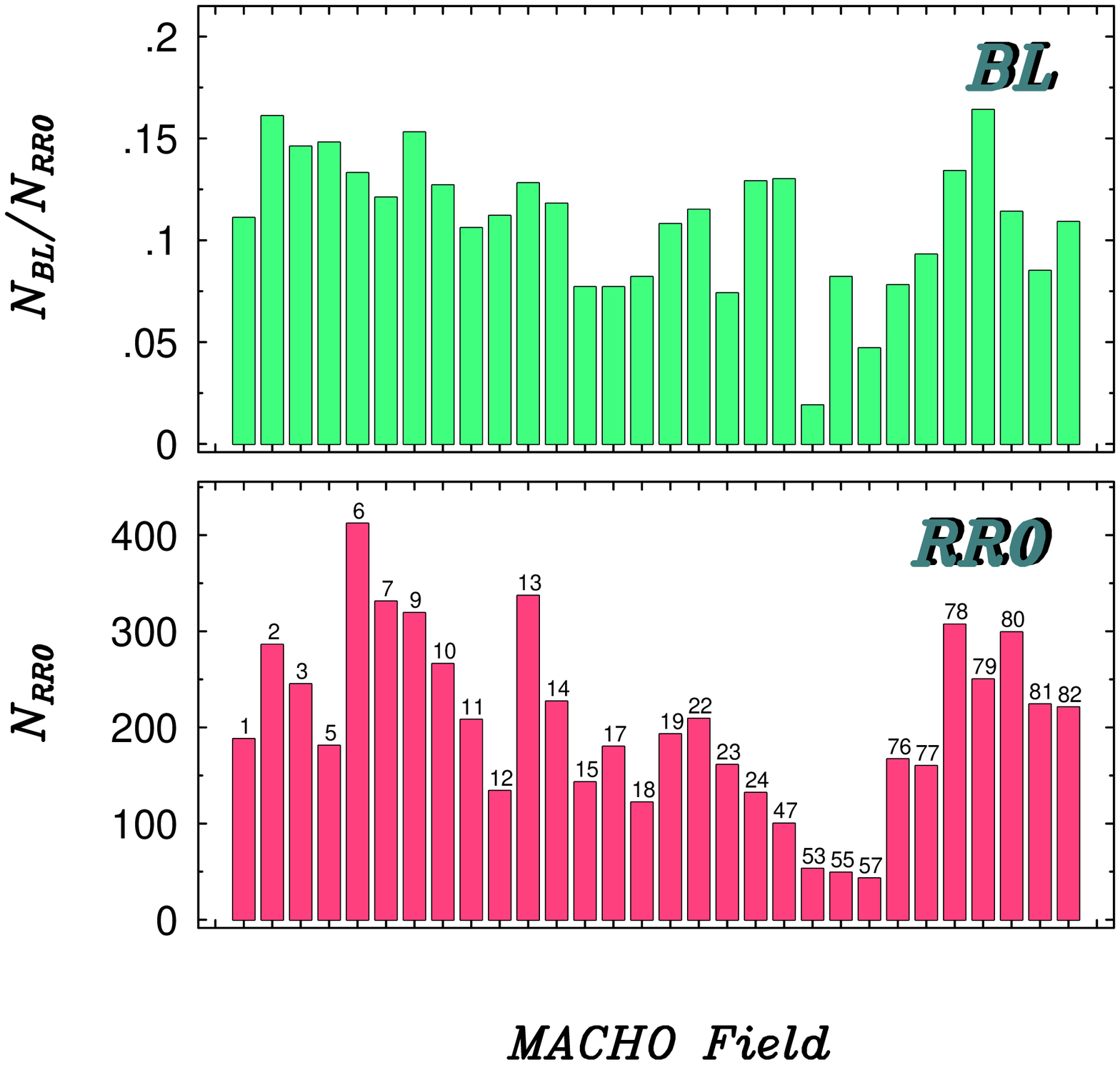,height=85mm,width=90mm}} 
\figcaption[h]{Field-by-field distribution of the number of all 
 fundamental mode RR~Lyrae stars (lower panel, with the field number 
 shown on the top of the columns) and the ratio of the Blazhko stars to 
 the total number of the fundamental mode RR~Lyrae stars (upper panel).}
\vskip 3mm 
We recall that the incidence rate of the RR1$-$BL stars in the LMC 
is 3.9\% (see A00). Although the frequency of the RR0$-$BL stars 
have proven to be lower than the commonly cited value of 20--30\%, 
they are still three times more common in the LMC than their RR1 
counterparts. It is possible that some RR1-BL stars might have 
been missed because of their low modulation amplitudes but it seems 
unlikely that so many BL stars were selectively lost due to the 
lower signal-to-noise ratio. 

%
%
\begin{deluxetable}{llllllll}
\tablecaption{Star-by-star comments on the frequency analysis}
\tablewidth{0pc}
\tablehead{MACHO ID & & & &TYPE & & & REMARK}
\startdata
1.3441.1031   & & & &   .     & & & \\
1.3442.1051   & & & &   .     & & & \\
1.3442.1107   & & & &   .     & & & \\
1.3442.1243   & & & &   BL2   & & & 
? Long mod. period, Df=0.00063 (1/T=0.00037) \\ 
1.3442.503    & & & &   .     & & & \\
1.3443.1313   & & & &   .     & & & \\
1.3443.1335   & & & &   .     & & & \\
1.3444.1187   & & & &   RR1   & & & \& PC \\
1.3444.56     & & & &   SOCEP & & & \\
1.3446.1877   & & & &   .     & & & \\
\enddata
\tablecomments{This table is available in its entirety only in the 
electronic version of this paper.}
\end{deluxetable}
%
%
%
\begin{deluxetable}{lccccccccccccccccc}
\tablecaption{Co-ordinates and main periods of the variables}
\tablewidth{0pc}
\tablehead{MACHO ID & & & & & $\alpha$(J2000.0) & & & & $\delta$(J2000.0) 
& & & & & & PERIOD [d]}
\startdata
1.3443.1335 & & & & & 05:01:26.0 & & & & $-$69:15:38 & & & & & & 0.4509916 \\
1.3444.1187 & & & & & 05:01:55.4 & & & & $-$69:11:31 & & & & & & 0.4239698 \\
1.3444.56   & & & & & 05:01:49.9 & & & & $-$69:12:46 & & & & & & 0.8057676 \\
1.3446.1877 & & & & & 05:01:42.7 & & & & $-$69:02:06 & & & & & & 0.5997655 \\
1.3446.1886 & & & & & 05:01:35.1 & & & & $-$69:02:06 & & & & & & 0.6061866 \\
1.3446.2004 & & & & & 05:01:37.9 & & & & $-$69:03:26 & & & & & & 0.4876528 \\
1.3449.1327 & & & & & 05:01:42.3 & & & & $-$68:52:06 & & & & & & 0.4839085 \\
\enddata
\tablecomments{This table is available in its entirety only in the 
electronic version of this paper.}
\end{deluxetable}
%
%
%
\begin{deluxetable}{rrrrrrrrrr}
\tablecaption{Fourier decompositions of the Blazhko variables}
\tablewidth{0pc}
\tablehead{
\multicolumn{1} {l} {MACHO ID} & \multicolumn {1} {c} {$f_0$}    & 
\multicolumn{1} {c} {$f_m$}      & \multicolumn {1} {c} {N}      & 
\multicolumn{1} {c} {$\overline {V}$} & \multicolumn {1} {c} {$\sigma$}  \\ 
\colhead{$a_1$} & \colhead{$a_2$} & \colhead{$a_3$} & \colhead{$a_4$} & 
\colhead{$a_5$} & \colhead{$a_6$} & \colhead{$a_7$} & \colhead{$a_8$} & 
\colhead{$a_9$} & \colhead{$a_{10}$} \\
\colhead{$\varphi_1$} & \colhead{$\varphi_2$} & \colhead{$\varphi_3$} & 
\colhead{$\varphi_4$} & \colhead{$\varphi_5$} & \colhead{$\varphi_6$} & 
\colhead{$\varphi_7$} & \colhead{$\varphi_8$} & \colhead{$\varphi_9$} & 
\colhead{$\varphi_{10}$}
}
\startdata
1.3446.2004  & 2.050641 & 0.011805 &  762 & 19.1238 & 0.0922 \\
0.0234 & 0.2982 & 0.0461 & 0.0246 & 0.1305 & 0.0557 & 0.0808 & 
0.0391 & 0.0240 & 0.0041 \\
3.7147 & 1.7243 & 4.0731 & 1.1233 & 5.6287 & 2.2739 & 3.5659 & 
1.6217 & 5.9261 & 2.8276 \\
1.3570.930   & 1.709248 & 0.008804 &  304 & 19.0981 & 0.0750 \\
0.0138 & 0.3227 & 0.0641 & 0.0284 & 0.1512 & 0.0255 & 0.1033 & 
0.0613 & 0.0365 & 0.0174 \\
5.5545 & 6.0614 & 0.1049 & 2.3508 & 1.9917 & 2.1388 & 4.3896 & 
0.4926 & 2.6433 & 5.2757 \\
\enddata
\tablecomments{This table is available in its entirety only in 
the electronic version of this paper.}
\end{deluxetable}
%
%
%
\begin{deluxetable}{lccc}
\tablecaption{Variables with multiple identification numbers}
\tablewidth{0pc}
\tablehead{MACHO ID & PERIOD [d] & $\alpha$(J2000.0) & $\delta$(J2000.0)}
\startdata
18.3446.4944    &   0.6061888 & 05:01:35.0 & $-$69:02:06 \\
1.3446.1886     &   0.6061866 & 05:01:35.1 & $-$69:02:06 \\ \\
1.3446.2004     &   0.4876528 & 05:01:37.9 & $-$69:03:26 \\
18.3446.4648    &   0.4876523 & 05:01:37.8 & $-$69:03:26 \\ \\
18.3449.4465    &   0.4939059 & 05:01:21.3 & $-$68:51:30 \\
1.3449.1459     &   0.4939031 & 05:01:21.3 & $-$68:51:31 \\
\enddata
\tablecomments{This table is available in its entirety only in the 
electronic version of this paper.}
\end{deluxetable}
%
%
%
\begin{deluxetable}{lrr}
\tablecaption{Average Fourier parameters of the singly-periodic RR0
and Blazhko stars}
\tablewidth{0pc}
\tablehead{Parameter & Average(RR0) & Average(BL)}
\startdata
$A_0$       &    $19.340\pm 0.008$  &  $19.351\pm 0.010$ \\
$A_1$       &    $ 0.293\pm 0.003$  &  $ 0.295\pm 0.003$ \\
$A_2$       &    $ 0.136\pm 0.001$  &  $ 0.123\pm 0.001$ \\
$A_3$       &    $ 0.094\pm 0.001$  &  $ 0.073\pm 0.001$ \\
$A_{tot}$   &    $ 0.843\pm 0.008$  &  $ 0.755\pm 0.007$ \\
$\phi_{21}$ &    $ 2.389\pm 0.005$  &  $ 2.365\pm 0.008$ \\
$\phi_{31}$ &    $ 5.076\pm 0.011$  &  $ 4.922\pm 0.014$ \\
$\phi_{41}$ &    $ 1.571\pm 0.019$  &  $ 1.301\pm 0.025$ \\
\enddata
\tablecomments{The symbol $A_{tot}$ denotes the total (peak-to-peak) 
amplitude without modulation. Errors show the standard deviations of 
the means as estimated from the scatter of the observed values. A 
$4\sigma$-cut is applied in the computation of the $A_0$ averages.}
\end{deluxetable}

Table~2 lists the identification numbers, classification and comments 
concerning all individual frequency spectra. Primary periods (those 
which appear with the highest amplitude in the frequency spectra of 
the original data) are listed in Table~3. Fourier decompositions of 
all BL stars are given in Table~4. Denoting the fundamental mode and 
Blazhko (or modulation) frequencies by $f_0$ and $f_m$, respectively, 
a Fourier sum with frequencies $\nu_1=f_0-f_m$, $\nu_2=f_0$, 
$\nu_3=f_0+f_m$, $\nu_4=2f_0-f_m$, $\nu_5=2f_0$, $\nu_6=2f_0+f_m$, 
$\nu_7=3f_0$, $\nu_8=4f_0$, $\nu_9=5f_0$, $\nu_{10}=6f_0$ has been 
fitted to each light curve and outliers were successively omitted 
until no data points deviated from the fitted curve by more than 
$3\sigma$ (i.e., three times the standard deviation of the residuals). 
The principle effect of these omissions is to remove photometry 
acquired in conditions of very poor seeing. Amplitudes and phases 
refer to the following type of decomposition: 
$V(t)=\overline {V} + a_1sin(2\pi(t-t_0)\nu_1+\varphi_1) + ...$. 
Here $t$ is given in HJD$-2400000.0$ and the epoch $t_0$ is set equal 
to zero for all variables. 

Modulation frequencies, as listed in Table~4, have been computed through 
the combination of the visual inspection of the frequency spectra and 
the automatic search for the ''strongest secondary component'' as 
described in Sect.~4. The visual inspection has resulted the revision 
of the $f_m$ values obtained by the automatic search for 19 variables. 
The failure of the automatic search in these cases was caused by various 
sources, mostly by large power residuals at the main pulsation component,  
but high noise level and aliasing were also matters of concern.  

Multiple identifications due to field overlaps are listed in Table~5. 
For full time series of all variables we refer to the {\sc macho} Project 
home page. Note that the time series available there are for the full 
span of {\sc macho} Project observations, unlike the data analyzed in 
this paper.

%
%
 
\section{COMPARISON OF THE GLOBAL PROPERTIES OF THE RR0 AND BL STARS}

The purpose of this section is to compare the overall photometric 
parameters of singly-periodic and Blazhko variables. We include all 
BL stars, because a preliminary investigation did not find any 
differences between the two subgroups with respect to the relations 
to be discussed in this section. 
 
%
%
%
\vskip 3mm
\centerline{\psfig{file=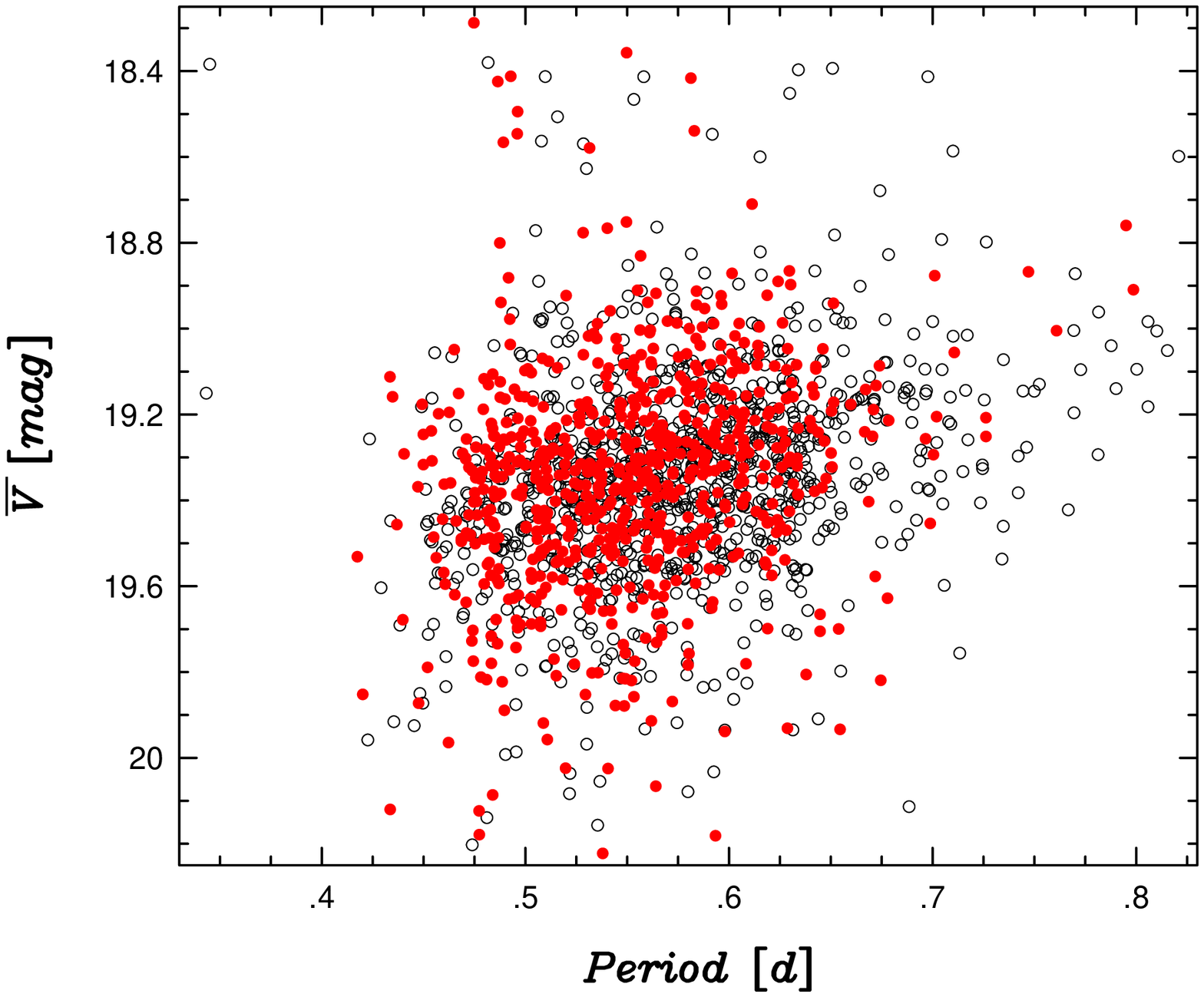,height=70mm,width=90mm}} 
\figcaption[h]{Comparison of the average magnitudes $\overline V$ 
for the singly-periodic RR0 stars (open circles) and for the RR0-BL stars 
(filled circles).}
\vskip 3mm 

First, we examine the average magnitudes. These values are computed 
from a Fourier fit described in the previous section. We used 10th-order 
fits for both the BL and singly-periodic variables. For the BL stars 
the fits include the basic pulsation component and its harmonics only 
up to order 6 because of the addition of the modulation components 
at $f_0$ and $2f_0$. We found that this fitting procedure gives 
reliable values for the low-order modulation and Fourier components. 
The average magnitudes are plotted in Figure~2. In order to avoid 
unnecessary crowding, we took a subsample of 1000 variables from 
{\sc macho} fields 1, 10, 11, 12, 13, 14 to represent the RR0-S stars. 

First of all, it is seen that both groups of stars cover the same 
ranges of parameters in this diagram. Based solely on their 
absolute magnitudes, Blazhko variables are indistinguishable 
from singly-periodic RR0 stars. Indeed, as shown in Table~6, the 
average of the mean magnitudes of both classes agree within their 
statistical errors. Both classes of stars display large scatter 
in $\overline {V}$. This is due to various effects, including 
crowding, inhomogeneous reddening within the LMC, differences in 
the absolute magnitudes, finite geometrical extension of the LMC 
and, of course, measurement/calibration errors. Despite the large 
scatter, some trend in the average magnitudes is observable. 
This trend is in qualitative agreement with the well-known relation 
between period and luminosity observed in other stellar clusters 
(e.g., Kov\'acs \& Walker 2001), and also expected from stellar 
evolution theory, which predicts brighter horizontal branches at 
lower metallicities (i.e., at longer periods -- see Castellani, 
Chieffi \& Pulone 1991).

%
%
%
\vskip 0mm
\centerline{\psfig{file=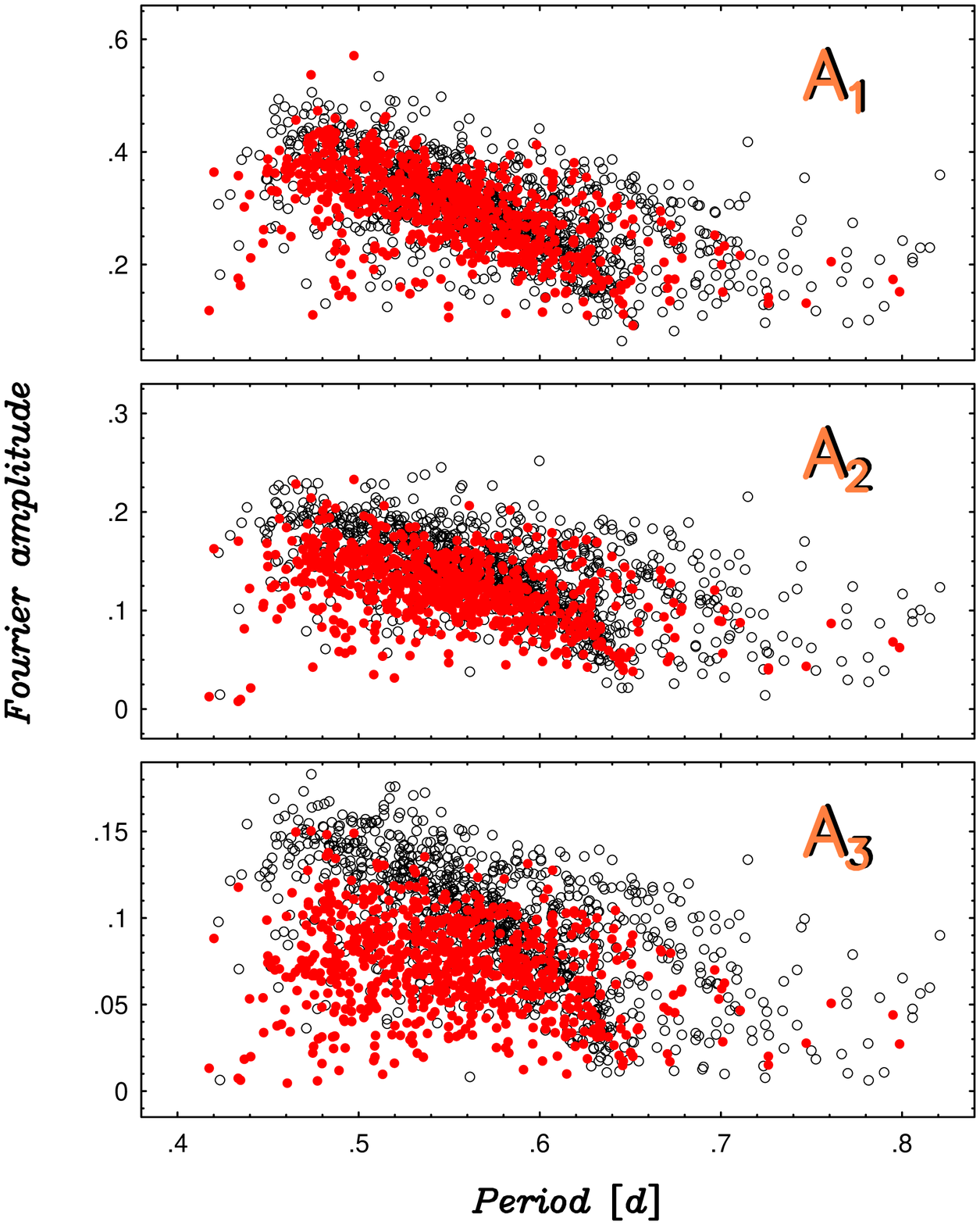,height=120mm,width=90mm}}
\vskip -3mm  
\figcaption[h]{Comparison of the Fourier amplitudes of the $f_0$, 
$2f_0$ and $3f_0$ components of the singly-periodic RR0 stars 
(open circles) and of the BL stars (filled circles).}
\vskip 3mm

Next we turn to the comparison of the Fourier parameters related 
to the main pulsation component $f_0$ and its harmonics. Denoting 
by $A_i$ the amplitude and by $\phi_i$ the phase of the Fourier 
component at frequency $if_0$, and employing the standard, 
epoch-independent phase difference $\phi_{i1}=\phi_i - i\phi_1$, 
Figures~3 and 4 show the variations of these quantities as functions 
of the period. It is seen that with increasing order, the Fourier 
amplitudes of the BL stars become significantly lower than those 
of the RR0-S stars (at the same period). A similar, but much weaker 
effect also exists among the phases as can be inferred from the 
computed average values and from their errors (see Table~6). This 
result implies that the modulation-free light curves of the BL 
stars show less skewness and have lower total amplitudes than 
their singly-periodic counterparts (see Table~6). The less 
significant difference between the phases suggests that the 
$(P_0, \phi_{31})\rightarrow$[Fe/H] relation derived by 
Jurcsik \& Kov\'acs (1996) may have a wider validity -- when 
applied to the average light curve -- than was originally suggested 
by these authors on the basis of the temporal Fourier decompositions 
of a few BL stars. (Lack of similarity between the temporal lightcurves 
of the BL stars and those of the single-mode RR0 stars has recently 
been studied by Jurcsik, Benk\H o \& Szeidl 2002.) It is also worth 
noting that the definite correlation between the period and the phases 
implies that the metallicity of the majority of the RR0 stars in the 
LMC is confined to a relatively narrow interval, as follows from the 
$P_0\rightarrow P_1/P_0$ diagram for the double-mode variables 
(Popielski, Dziembowski \& Cassisi 2000; Kov\'acs 2001), and also 
from direct spectroscopic measurements (Bragaglia et al. 2001). 

%
%
\vskip 2mm
\centerline{\psfig{file=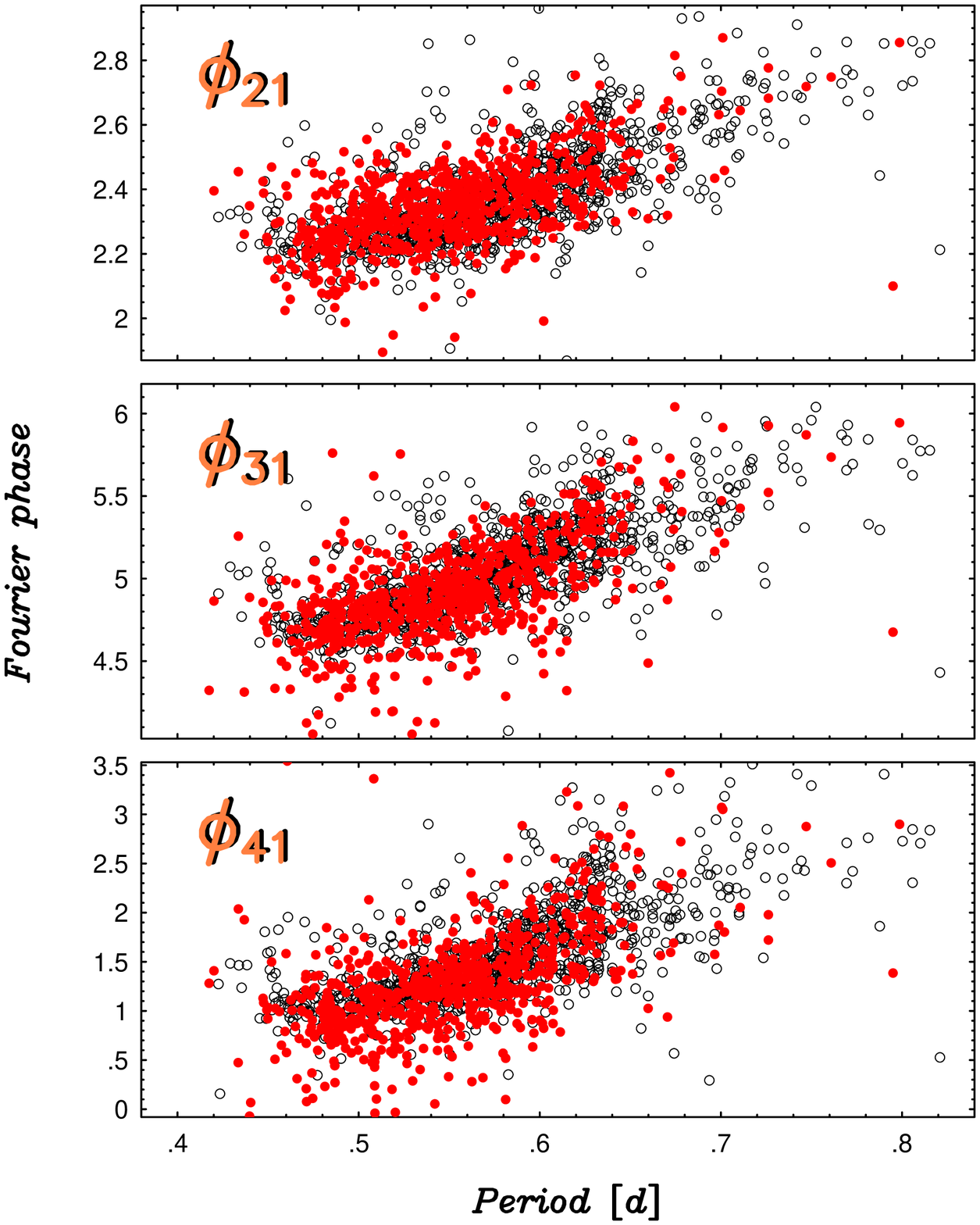,height=120mm,width=90mm}} 
\vskip -3mm 
\figcaption[h]{Comparison of the relative Fourier phases of the 
$2f_0$, $3f_0$ and $4f_0$ components of the singly-periodic RR0 
stars (open circles) and of the RR0-BL stars (filled circles).}
%

%
%

\section{MODULATION FREQUENCY AND AMPLITUDE DISTRIBUTIONS}

In this section we will discuss both the frequency with which
modulation timescales appear in the BL population and the incidence
with which the stronger modulation component is at higher or lower
frequency than the primary pulsation peak. For this purpose, we define 
the sign of the modulation frequency $f_m$ 
($\vert f_m \vert \equiv \vert f_0-f_{\pm} \vert$) to be positive 
if $A_+ > A_-$ and negative if the reverse is true (here $A_+$ and 
$A_-$ are, respectively, the amplitudes of the modulation components 
on the higher and lower frequency sides of the main pulsation component). 
With the above definition we can include all BL stars and study if 
there is a preference for a dominant frequency/amplitude pattern. 
Figure~5 shows the resulting distribution function. It is clearly seen 
that there is a strong preference for frequency patterns where
the modulation amplitude on the higher frequency side is larger than 
its lower frequency counterpart. We find that 74\% of all BL stars have 
positive $f_m$. Similar ratios are observed when the two types of BL 
stars are studied separately: $f_m>0$ for 79\% of the BL1 stars 
and 68\% of the BL2 stars. Blazhko variables with very long modulation 
periods and those with periods as short as 50 days occur with about the 
same probability. For shorter modulation periods the number of BL stars 
sharply decreases and reaches very small incidence rates for modulation 
periods shorter than about 25~d. Nevertheless, one can still find a few 
stars with very short Blazhko periods: 9.5360.804 and 6.6212.1121 have 
$f_m=0.050$ and $0.122$~d$^{-1}$, respectively, with the latter value 
being the only one in the whole sample exceeding $0.1$~d$^{-1}$. We 
also note that similarly short modulation period was observed in AH~Cam 
by Smith et al. (1994).  
 
%
%
\vskip 3mm
\centerline{\psfig{file=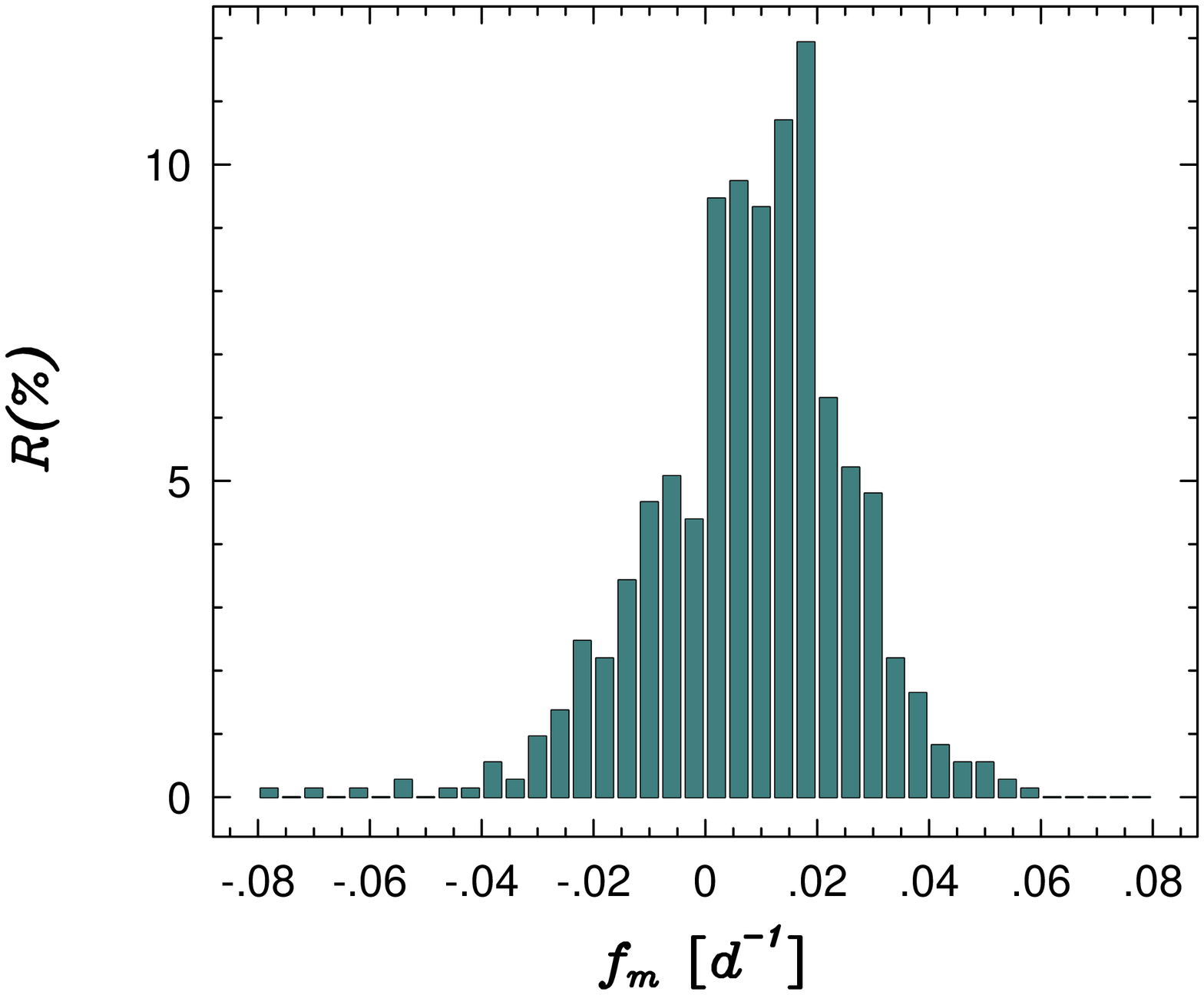,height=75mm,width=85mm}} 
\figcaption[h]{Distribution of the modulation frequencies for the 
731 BL stars. See text for the sign convention of $f_m$.}
\vskip 3mm 

It is also a matter of interest whether the small deviations from a 
strictly equidistant frequency spacing in the case of the BL2 stars 
are statistically significant. We recall that in the course of a similar 
analysis of the RR1 stars it was shown that the observed deviations are 
not significant (see A00). Following the same notation as in A00, we 
compute the quantity $\delta f \equiv f_{+}+f_{-}-2f_0$ to characterize 
the degree of deviation from equidistant spacing. We show the distribution 
function obtained from the analysis of the BL2 stars in Figure~6. 
We select the triplets from the frequency spectra trough the following 
automated procedure:
\begin{itemize}
\item
Compute successively pre-whitened amplitude spectra in the 
$\pm 0.1$~d$^{-1}$ neighborhood of the main pulsation component. 
Five  pre-whitening cycles are computed and the highest peak  
obtained after each cycle is stored. 
\item
The ``strongest secondary component'' $f$ is chosen from the stored peak 
frequencies by requiring $\vert f - f_0 \vert > c_1/T$, where $c_1=1.5$, 
$T$ is the total timespan of the respective dataset. This condition 
produces a distinction between PC and long-period BL stars.
\item
A third peak is selected from frequencies appearing on the side opposite 
to the main component where the ``strongest secondary component'' was 
found. The variable is classified as a BL2-type if 
$\vert \delta f\vert < c_2/T$, where $c_2=3.0$. This condition is 
used to select triplets with ``approximately'' equidistant frequency 
spacings.    
\end{itemize}     
In nine cases out of the 331 BL2 variables, this automatic 
selection method could not find a triplet, usually because the 
large PC component coexisting with the triplet did not allow the 
process to reach the level of the Blazhko components within the 
five pre-whitening cycles. A high noise level was also a factor 
in some cases. Therefore, Figure~6 shows the results for the 
remaining 322 stars. (It is noted that five variables from this 
sample yielded considerably smaller $f_m$ values than the ones 
listed in Table~4. Again, this difference is accounted for by 
the large PC component in these stars. In order to maintain a   
close compatibility with the subsequent statistical tests, we 
keep these different modulation frequencies in the investigation 
of the distribution of the non-equidistant spacings.)  

We see that, with few exceptions, all frequency deviations are 
within $\pm 0.0002$~d$^{-1}$. This is about one half of the 
characteristic linewidth (i.e., $1/T\approx 0.00037$~d$^{-1}$). 
In order to assess if this degree of unequal spacing still can be 
explained as a result of noise-induced frequency-shift, we perform 
the following test. Artificial time-series are generated with 
equidistant triplets given by the Fourier decompositions (see Sect.~2) 
of the 331 BL2 stars. Then, a Gaussian white noise is added to these 
data with the standard deviation of the fit of the original data. 
In the next step these test data are processed in the same way as 
the real observations. Finally, we use the peak frequencies obtained 
from the analysis to compute $\delta f$, and plot the corresponding 
histogram. The result is shown in Figure~7. Comparison with Figure~6 
reveals that the two distributions are very similar to each other, 
although the test data show a smaller dispersion and no outliers 
above $\vert \delta f \vert > 0.0002$~d$^{-1}$. The derived standard 
deviations are consistent with this conclusion. After applying a 
$3\sigma$ criterion for the omission of the outliers, we find 
$\sigma_{\delta f}^{obs}=6.3\times 10^{-5}$~d$^{-1}$ and          
$\sigma_{\delta f}^{test}=4.9\times 10^{-5}$~d$^{-1}$. By performing 
the above test for 10 different realizations of the noise for each 
time-series, we get $\sigma_{\delta f}^{test}=(4.59\pm 0.26)\times 10^{-5}$~d$^{-1}$, where the error represents the standard deviation 
of the ten $\sigma_{\delta f}^{test}$ values obtained with the various realizations. We conclude that the difference between the observed and 
test values is statistically significant. Therefore, the observed 
small deviations from equidistant spacing cannot be accounted for 
by purely noise-induced frequency-shifts. 

In checking the notes (see Table~2) related to the nine variables 
mentioned before for not passing the automatic selection criterion 
and those 18 stars which did not pass the $3\sigma$ criterion with 
respect to the symmetric spacing, we find the following. Except for 
two cases, the lightcurves of the variables have some peculiarity -- 
often of a PC nature -- but we also find remarks on high noise level 
and small second modulation component. In one case there is no comment 
and in a single case the observed frequency is declared `nice'. 

%
%
%
\vskip 3mm
\centerline{\psfig{file=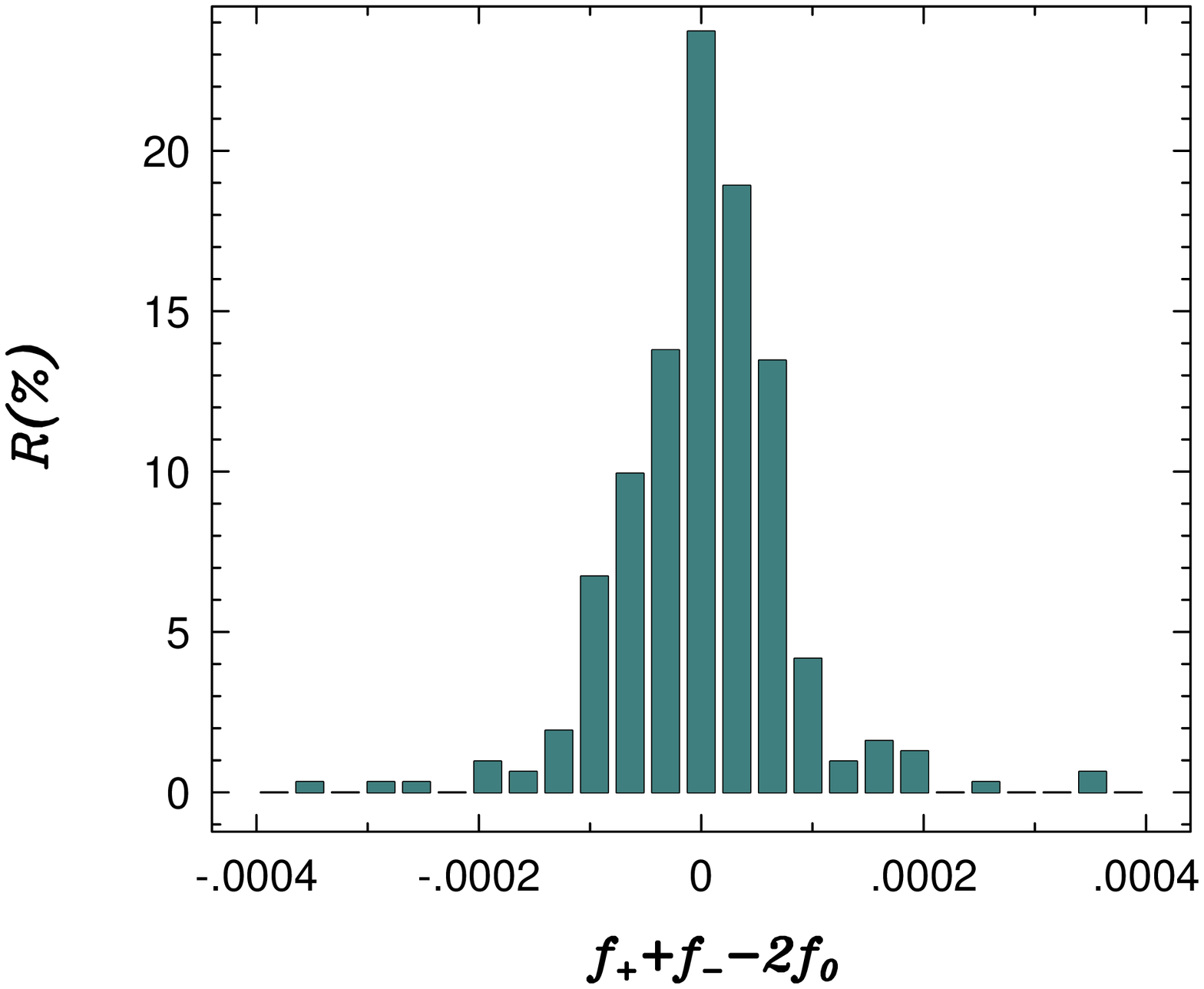,height=75mm,width=85mm}} 
\figcaption[h]{Distribution of the differences in the triplet spacings 
for the BL2 stars.}
\vskip 3mm 
%

%
%
\vskip 3mm
\centerline{\psfig{file=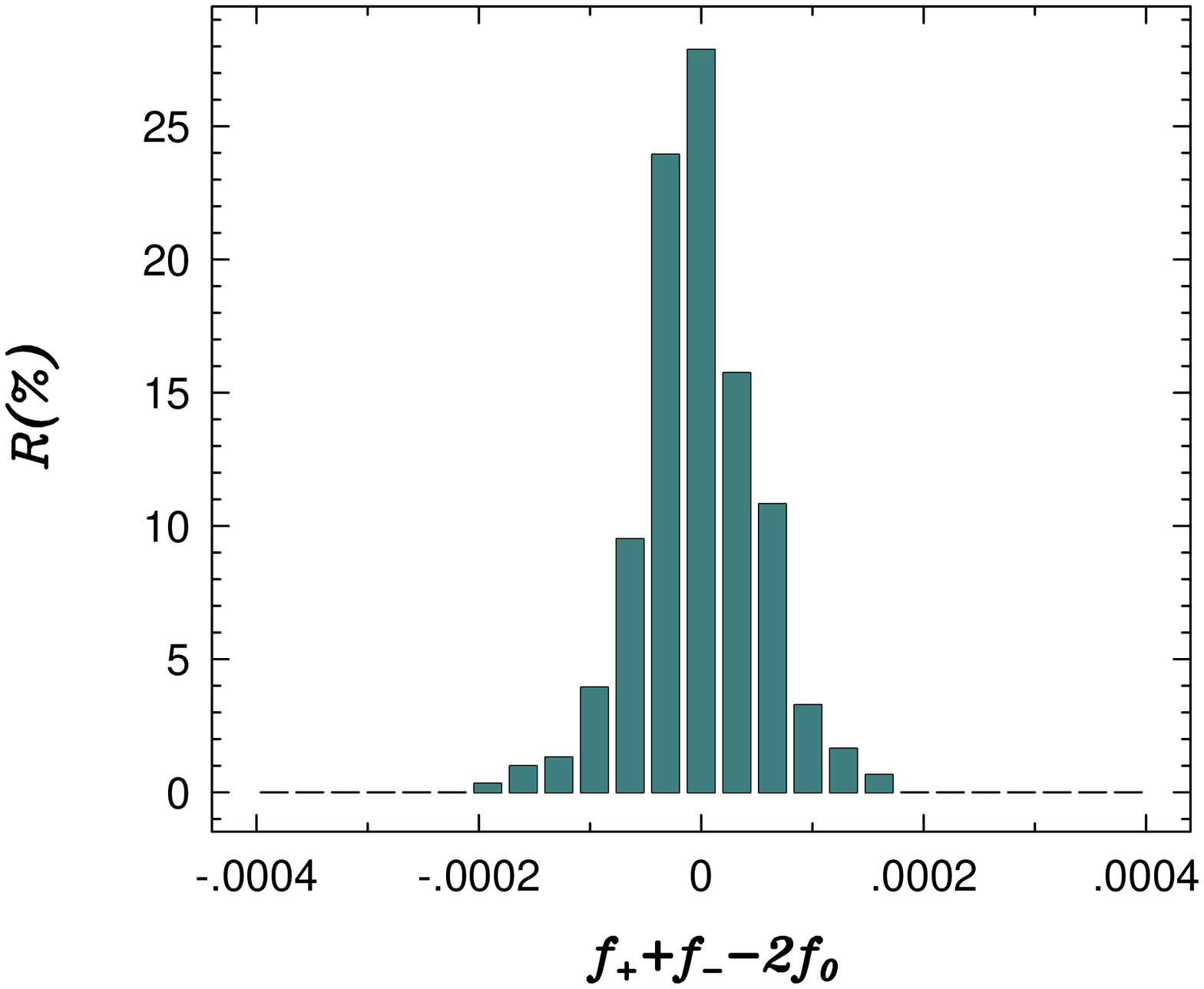,height=75mm,width=85mm}} 
\figcaption[h]{Distribution of the differences in the triplet spacings 
obtained from {\bf test data} as described in the text.}
\vskip 3mm 

Our conclusion from all these tests is that the cause of non-equidistant 
frequency spacing has various origins, including cases when neither 
the noise level, nor peculiar additional features are suspected to 
be responsible for the deviation. The cause of the non-equidistant 
spacing should be studied in detail on a star-by-star basis by 
considering various effects in addition to the noise-induced 
frequency-shift as discussed above.  

In some models a resonance between the fundamental and Blazhko periods 
$P_{BL}=1/f_m$ could be a matter of interest. Therefore, we compute 
the fractional part of $P_{BL}/P_0$ and plot its distribution function. 
As can be verified from Figure~8, there are no resonances between the 
two periods and the distribution is essentially flat for all fractional 
values. The fluctuations are of statistical origin, because the 
relative standard deviation in an average bin is 23\% (assuming a 
Poisson distribution).
 
%
%
%
\vskip 3mm
\centerline{\psfig{file=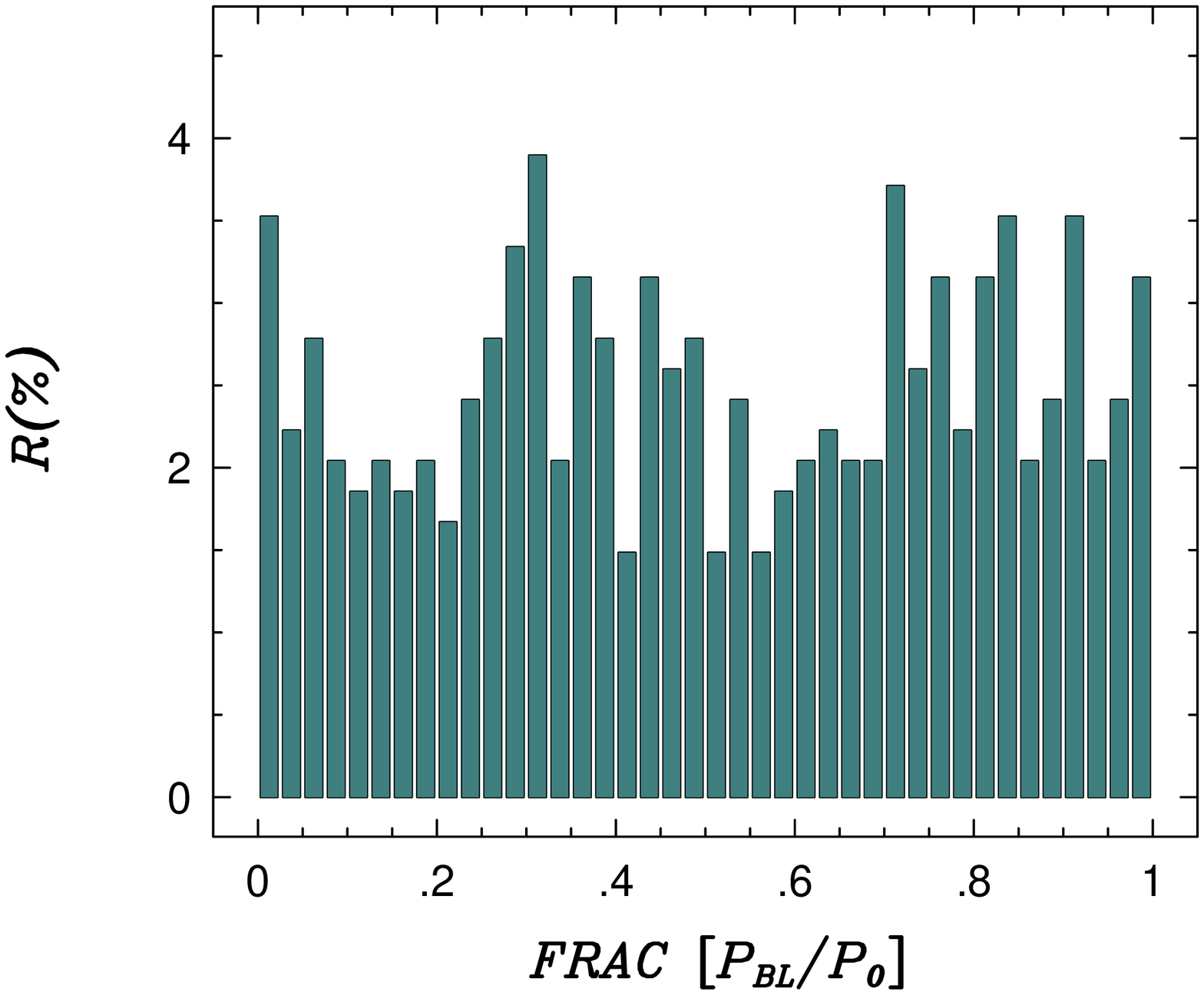,height=75mm,width=85mm}} 
\figcaption[h]{Distribution of the fractional part of $P_{BL}/P_0$ 
for the 731 BL stars.}
\vskip 3mm 
%
%

%
%
\vskip 3mm
\centerline{\psfig{file=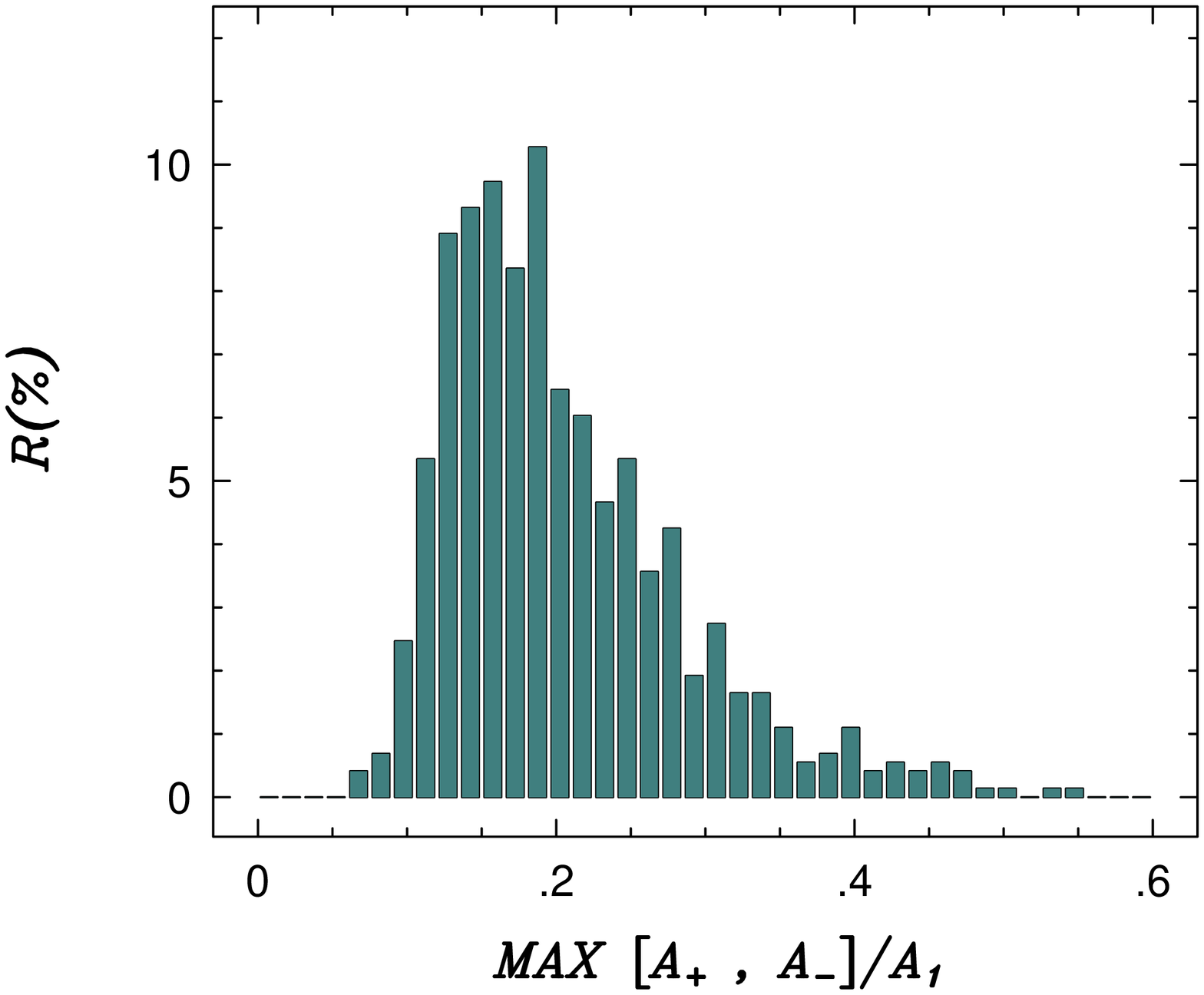,height=75mm,width=85mm}} 
\figcaption[h]{Distribution of the ratio of the maximum modulation 
amplitude to the main pulsation amplitude for the 731 BL stars.}
\vskip 3mm 

We now investigate the distribution of the modulation amplitudes. 
There are two important properties to address: (a) range of the 
relative (pulsation vs. modulation) strength of the modulation 
amplitudes, and (b) degree of asymmetry of the left- ($A_{-}$) and 
right- ($A_{+}$) modulation components. We employ the Fourier 
decompositions both for the BL1 and for the BL2 stars to get the 
observed values of $A_1$, $A_{-}$ and $A_{+}$. Since noise always 
introduces some non-zero excess power at all frequencies, a 
distortion of the distributions at low modulation amplitudes is 
expected. 
  
%
%
\vskip 3mm
\centerline{\psfig{file=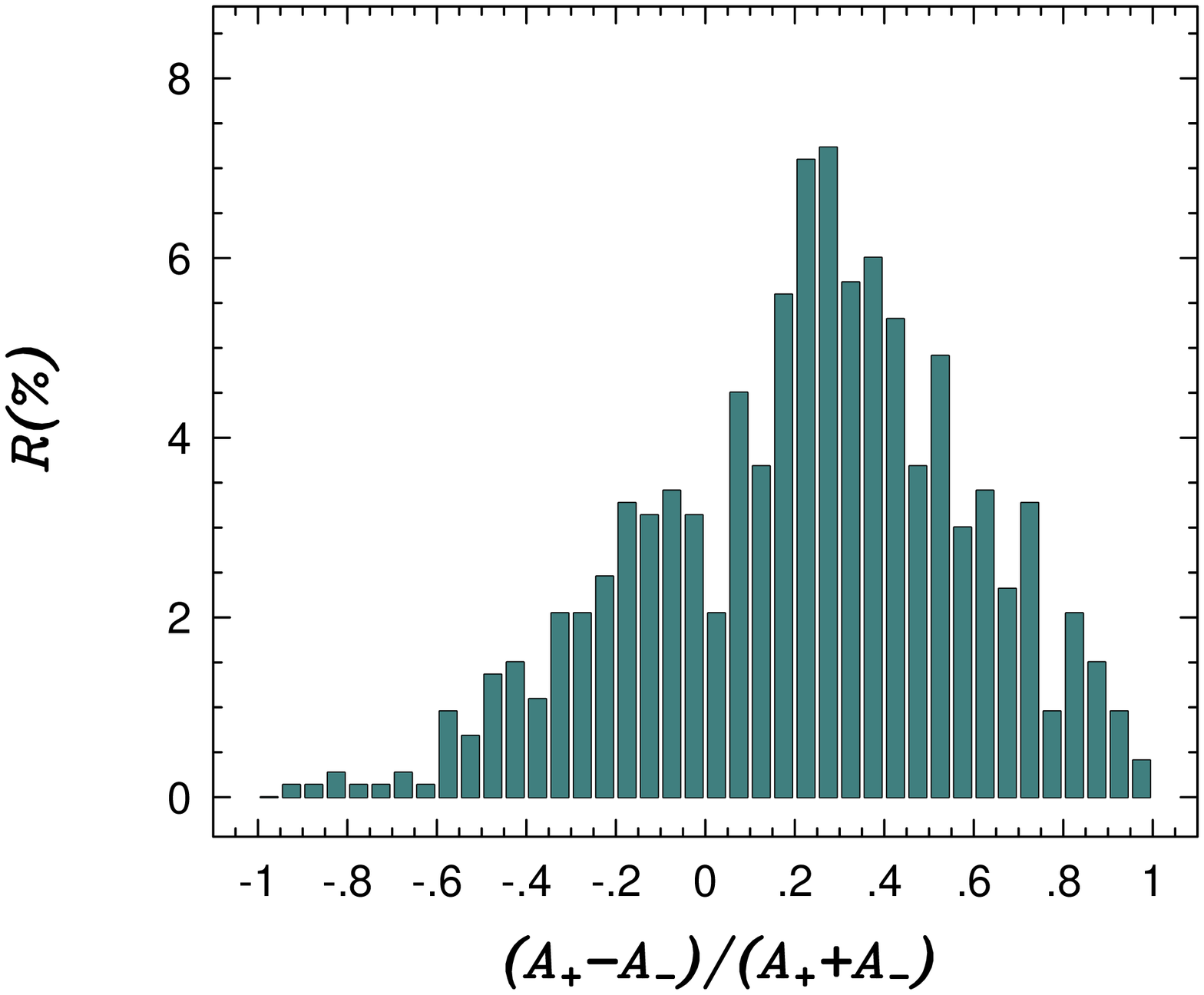,height=75mm,width=85mm}} 
\figcaption[h]{Distribution of the asymmetry ratio for the 731 BL stars.}
\vskip 3mm 

We plot the distribution of the relative modulation strength in 
Figure~9. In order to minimize the effect of noise mentioned 
above, we take the largest modulation component in each case. 
The distribution sharply peaks at $\approx 0.15$ and decays with 
a long tail after $\approx 0.2$. It is important to observe that 
there is a wide range of possible modulation strengths from very 
weak to very strong with the largest having $A_{\pm}/A_1>0.4$. 
Considering extreme modulation strengths, we call attention to 
the specific cases of variables 2.5870.4967 with 
{\sc max(}$A_{\pm}/A_1$)$=0.55$ and 9.4997.754, with 
{\sc max(}$A_{\pm}/A_1$)$=0.10$. The modulation 
components are clearly detected for both stars.

The asymmetry in the modulation amplitudes is a significant problem 
for both presently available models (Shibahashi 2000, Nowakowski 
\& Dziembowski 2001). To quantitize the degree of asymmetry, we 
introduce the following parameter
 
%
%
\begin{eqnarray}
Q = {{A_{+}-A_{-}}\over {A_{+}+A_{-}}} \hskip 2mm .
\end{eqnarray}

We recall that the amplitudes are computed from Fourier fits which 
include both the $A_{-}$ and $A_{+}$ components. Q is found to
cover almost all values, partly as a result of noise and partly 
due to the presence of `hidden' BL2 variables among the BL1 stars. 
In fact, if we perform an automated search for symmetric frequency 
patterns in the way mentioned at the beginning of this section, we 
find that 160 out of the 400 BL1 stars could be classified as BL2-type. 
  
Figure~10 displays the distribution function of all BL stars. 
As we have already seen in Figure~5, there is a strong dominance 
of variables with $A_{+}>A_{-}$. The distribution is very broad 
but the most probable cases are those with $Q\approx 0.1$--$0.6$. 
Of course, when the above distribution is plotted only for the BL1 
variables, we find a deficit at small $\vert Q\vert$ values and 
the distribution is double-peaked at larger/smaller $Q$ values. 
The BL2 variables are mostly confined to the $Q=(-0.2,0.4)$ interval, 
but some extreme asymmetric cases can also be found. For example, 
variable 77.7793.1432 shows a large positive asymmetry with $Q=0.823$. 

%
%
 
\section{A SEARCH FOR ADDITIONAL MODULATION COMPONENTS BY SPECTRUM-AVERAGING}

Our standard method of analysis described in Sect.~2 did not 
find any definite signature of two interesting features (both of 
these are expected within the frameworks of either the oblique rotator 
or of the resonant mode coupling models). The first feature we 
expect is due to the long-term periodic variation of the average 
brightness. The presence of this component is expected from 
non-linear mode-coupling. The other possible feature is an
equidistant quintuplet structure around $f_0$. This property is 
predicted by the oblique rotator model and is also expected when 
mode-coupling occurs. Since such features are not visible in 
the observed (individual) frequency spectra, we employ a method 
of spectrum averaging in order to amplify such small signals 
(if they are present).   

Our procedure is the following. The huge differences in the lengths 
of the Blazhko periods require us to normalize the spectra in both 
the amplitude and frequency domains before averaging.  Therefore, 
after pre-whitening by the `trivial' components (i.e., $f_0-f_m$, 
$f_0$, $f_0+f_m$, etc.) and taking the Fourier transform of the 
residuals, each spectrum is divided by its variance for more effective 
noise filtering of the different quality spectra. (We always use 
amplitude spectra -- both in the computation of the individual and 
of the average spectra.) Then, all peaks are identified and shifted 
by the amount necessary to position the modulation components in 
the same place in each case. Then, this unified frequency band is 
divided into equidistant bins and the maximum amplitude values of 
the peaks are found for each bin. These values are used in the 
binned-and-normalized version of original frequency spectrum. 
Finally, all these spectra are added together and the average is 
plotted for visual inspection. To demonstrate the effectiveness 
of the method, Figure~11 reveals the result obtained by the analysis 
of a small sample of BL2 variables. 

We see that the signal-to-noise ratio of the average spectrum has 
dramatically improved compared to the majority of the individual 
spectra. The pre-whitening by the $f_0$ component (and its harmonics), 
results in a dip at $\Delta f/f_{mod}=0$. As we shall see in the 
following subsections, dips related to pre-whitening will show 
up more obviously when using a large set of variables. Occasionally, 
a rather large separation is observable between the peaks in the 
individual normalized spectra (e.g., variable 5.4288.4400). This 
effect is due to the long modulation periods of those stars, which 
results in a greater stretch in the frequency domain. Therefore, 
because of the finite temporal baseline of the observations, we 
observe fewer peaks in the very narrow frequency band corresponding 
to the long modulation period. PC behavior is also observable in 
some cases (e.g., {\sc macho} ID 5.4285.302). The variable 5.4891.855 
represents one of the handful cases where a quadruple structure is 
observed, but with unequal frequency spacings.  

%
%
%
\vskip 3mm
\centerline{\psfig{file=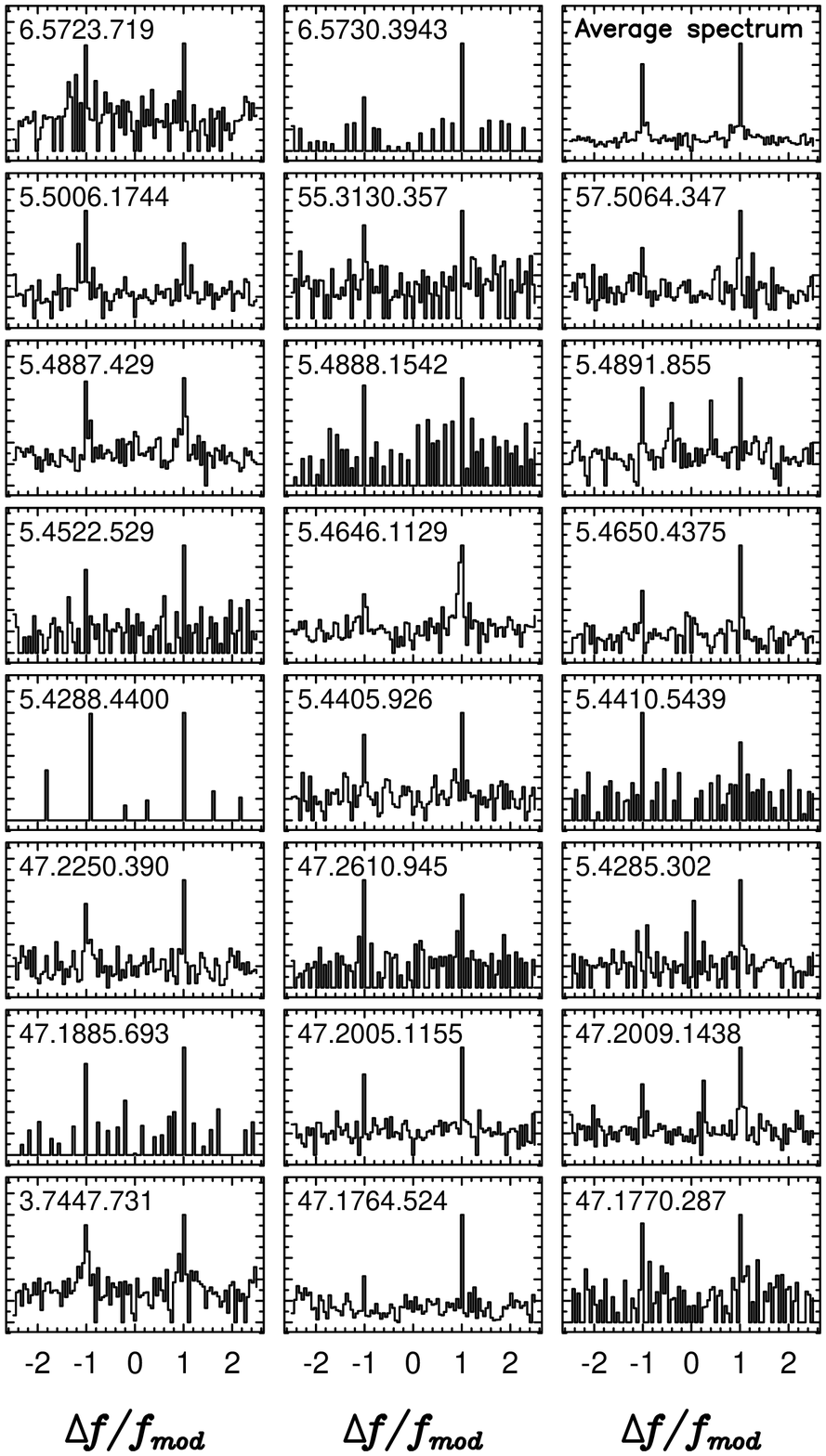,height=135mm,width=90mm}} 
\figcaption[h]{An example of spectrum-averaging. Individual 
normalized spectra are shown in each panel, except for the upper 
right one, where the resulting average spectrum is displayed. 
Each spectrum has been computed after pre-whitening by $f_0$ 
and its harmonics. Only the peak amplitudes are displayed. The 
symbol $\Delta f$ denotes $f-f_0$, where $f$ is the test frequency. 
All variables are classified as BL2.}
\vskip 3mm 
%

%
%

\subsection{Average light level}

In order to reach the highest possible signal-to-noise ratio, we 
employ all BL stars in the computation of an average spectrum. 
Individual frequency spectra are computed in the $[0,0.2]$~d$^{-1}$ 
band. By using $200$ bins in all cases, each spectrum is transformed 
in the frequency domain of $0 < \Delta f/f_m < 3$, where $\Delta f$ 
denotes the test frequency $f$. The result is somewhat sensitive to 
the choice of number of bins. If the bins are too wide (compared to 
the linewidth), then the hidden signal will be less effectively 
recovered because of the neighboring smaller peaks (which come from 
the noise) will lower the amount of power collected in the bin of 
interest. On the other hand, bins which are too narrow will result 
in a division of signal power into several adjacent bins due to 
the noise-induced frequency-shift of the peak associated with the 
signal. With the above choice of bin number we cover approximately 
the characteristic linewidth at the moderately short Blazhko period 
of $\approx 50$~days. This bin width is also appropriate for possible 
frequency shifts due to noise (see Figures~6 and 7). In order to avoid 
too large a deformation in the plot due to the large dip at zero 
frequency (each time series has zero average), the final spectrum 
is shifted downward by an appropriate amount, then stretched in the 
vertical axis to cover the $[0,1]$ interval by the [min,max] range 
of the spectrum.    

%
%
%
\vskip 3mm
\centerline{\psfig{file=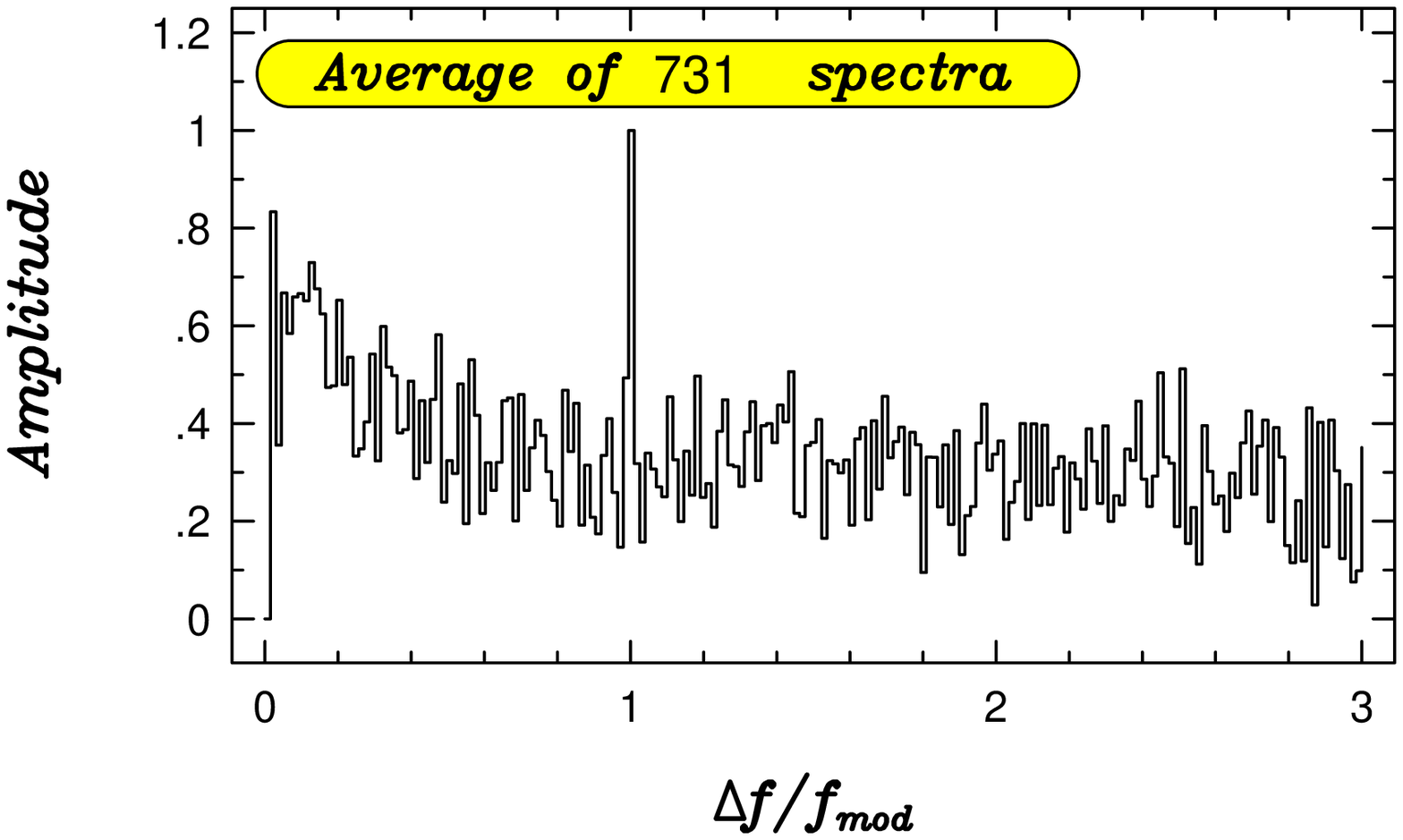,height=50mm,width=90mm}} 
\figcaption[h]{Average frequency spectrum of all BL stars in the 
proximity of the expected long-term modulation with the Blazhko 
frequency $f_{mod}$. The symbol $\Delta f$ denotes the test 
frequency $f$. Amplitudes and relative frequencies are normalized 
as described in the text.}
\vskip 3mm 
%

%
%
\vskip 3mm
\centerline{\psfig{file=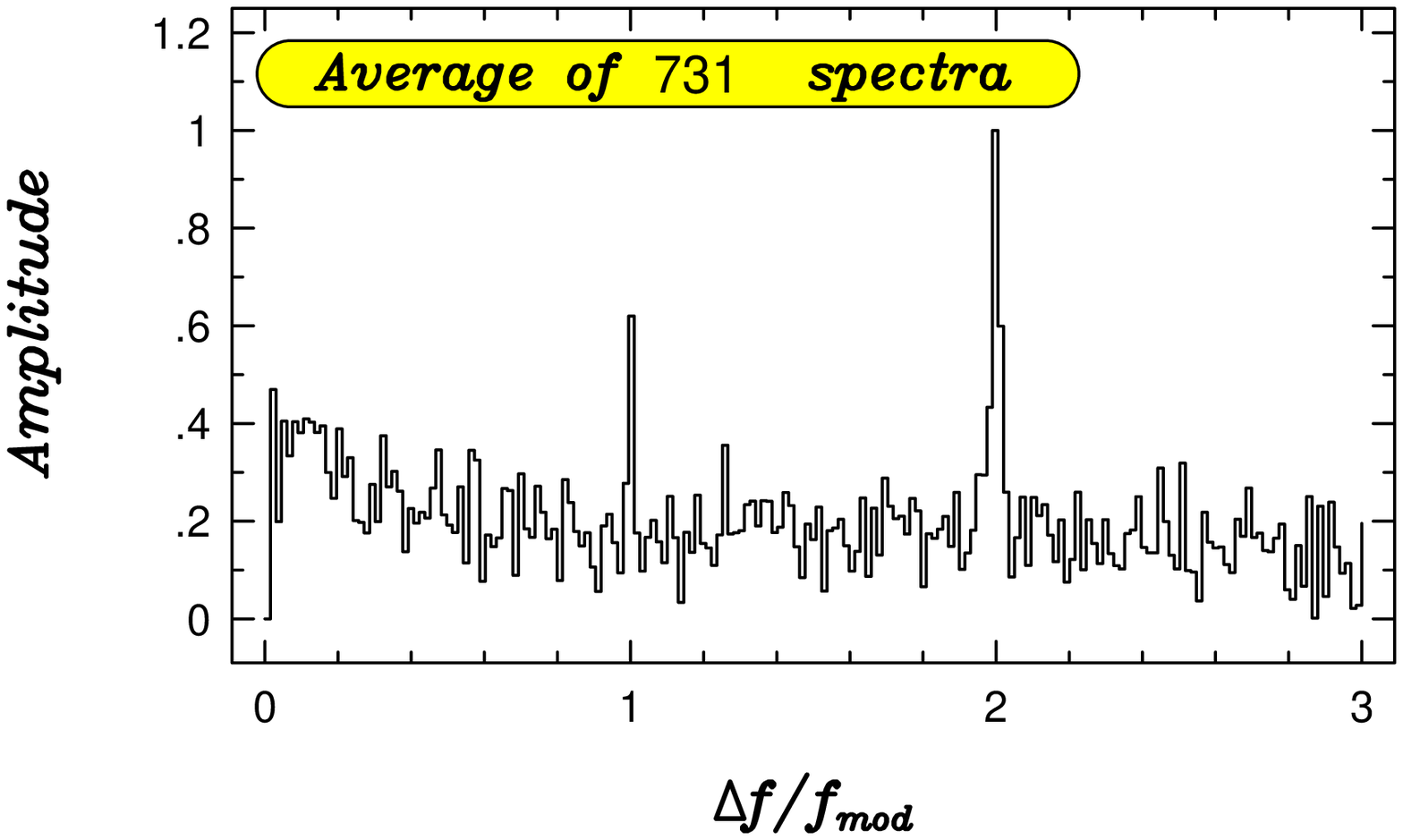,height=50mm,width=90mm}} 
\figcaption[h]{As in Figure~12 but a {\bf sinusoidal tracer} 
with an amplitude of $0.010$~mag and frequency $2f_{mod}$ added 
to the signals.}
\vskip 3mm 

Figure~12 shows the resulting amplitude spectrum. We observe a definite 
peak at $\Delta f/f_{mod}=1.0$. Some earlier studies on Galactic Blazhko 
variables also indicated the presence of this component (Borkowski 1980;  
Kov\'acs 1995; Nagy 1998). However, the data used in those studies 
suffered from very strong aliasing by the main periods of the variables
caused by the observational technique employed when those variables 
were observed. As a result, the possibility of power leakage from the 
$f_0\pm f_m$ components to frequencies differing by $f_0$ from these 
modulation components is expected.

Since the observed amplitude is influenced by the size of bins and 
noise level, the simplest way to get an estimate on the amplitude and 
on the significance of the peak is to add a sinusoidal tracer of a 
given frequency to the signal. This has been done, and the result 
shown in Figure~13 is obtained. By comparing the two peaks we estimate 
the overall amplitude of the observed modulation in the average light 
level of about $0.006$~mag. 

A further characteristic feature observable in the spectrum is the 
slow rise of the signal power from $\Delta f/f_{mod}\approx 1.0$ to 
zero. This mild trend is most probably due to long-term instrumental 
or photometric calibration effects. 

%
%

\subsection{Quintuplet structure}

Side peaks at $f=f_0\pm 2f_m$ have been searched for in the same manner 
as the $f_m$ component. Here the number of bins used was $400$ because 
of the larger bandwidth required by this analysis. The average 
spectrum of all the BL stars is shown in Figure~14. In addition to 
some excess power very close to the prewhitened $f_0$ component 
and to $f_0\pm f_m$, we may also suspect a peak at $f_0-2f_m$, very 
close to the noise level. The excess power near $f_0$ (and probably 
also at $f_0\pm f_m$) is associated with the observed PC character 
of some variables. On average, this is a small effect, but in a few 
individual cases this may be comparable to the Blazhko effect (e.g. 
6.6570.771 and 78.5615.905).

%
%
%
\vskip 3mm
\centerline{\psfig{file=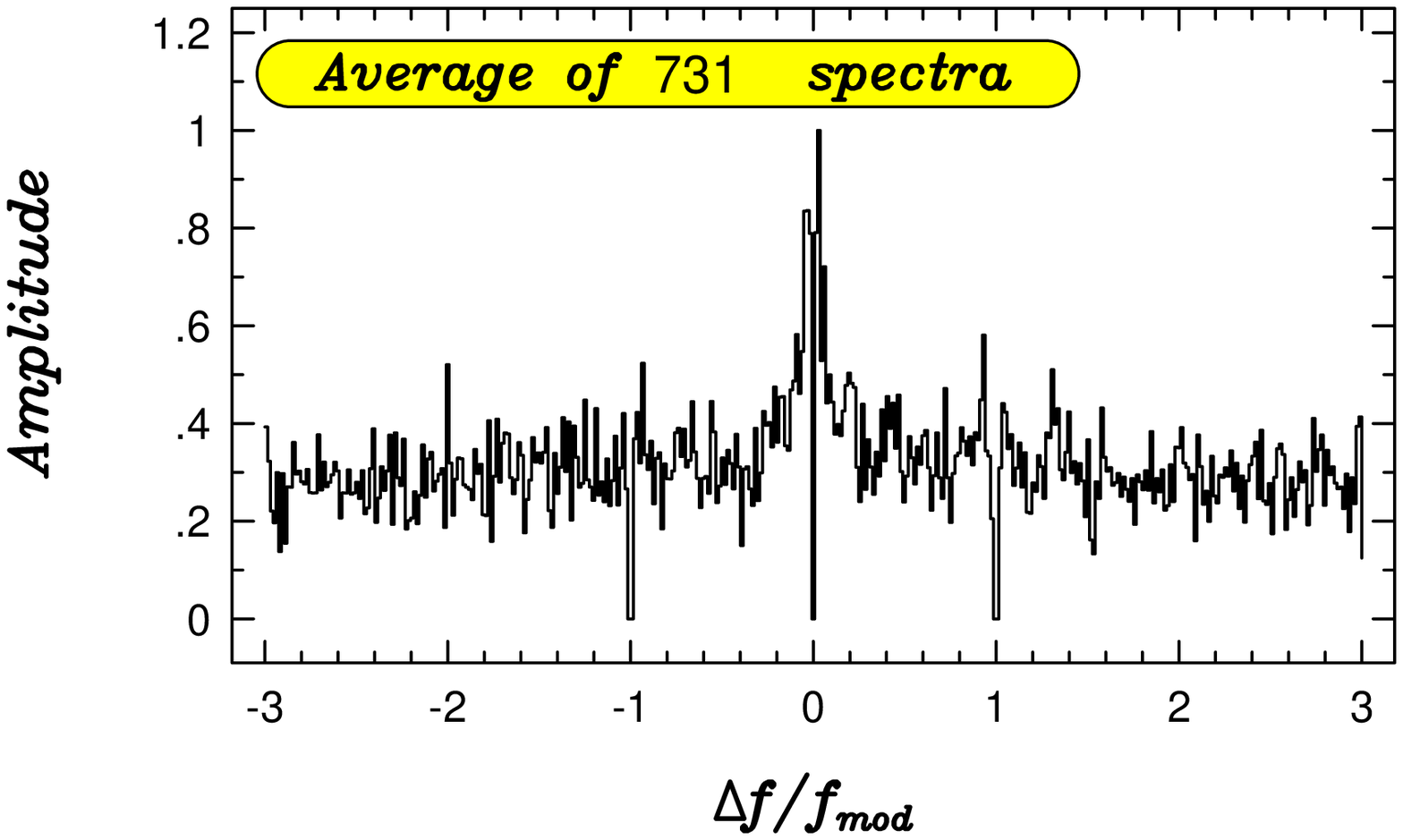,height=50mm,width=90mm}} 
\figcaption[h]{Average frequency spectrum of all BL stars in the 
neighborhood of $f_0$. The main pulsation ($f_0$) and modulation 
($f_{-}$ and $f_{+}$) components have been prewhitened before 
Fourier transformation. The symbol $\Delta f$ denotes $f-f_0$, 
where $f$ is the test frequency. Amplitudes and relative frequencies 
are normalized using a procedure described in the text.}
\vskip 3mm 
%
%
%
%
\vskip 3mm
\centerline{\psfig{file=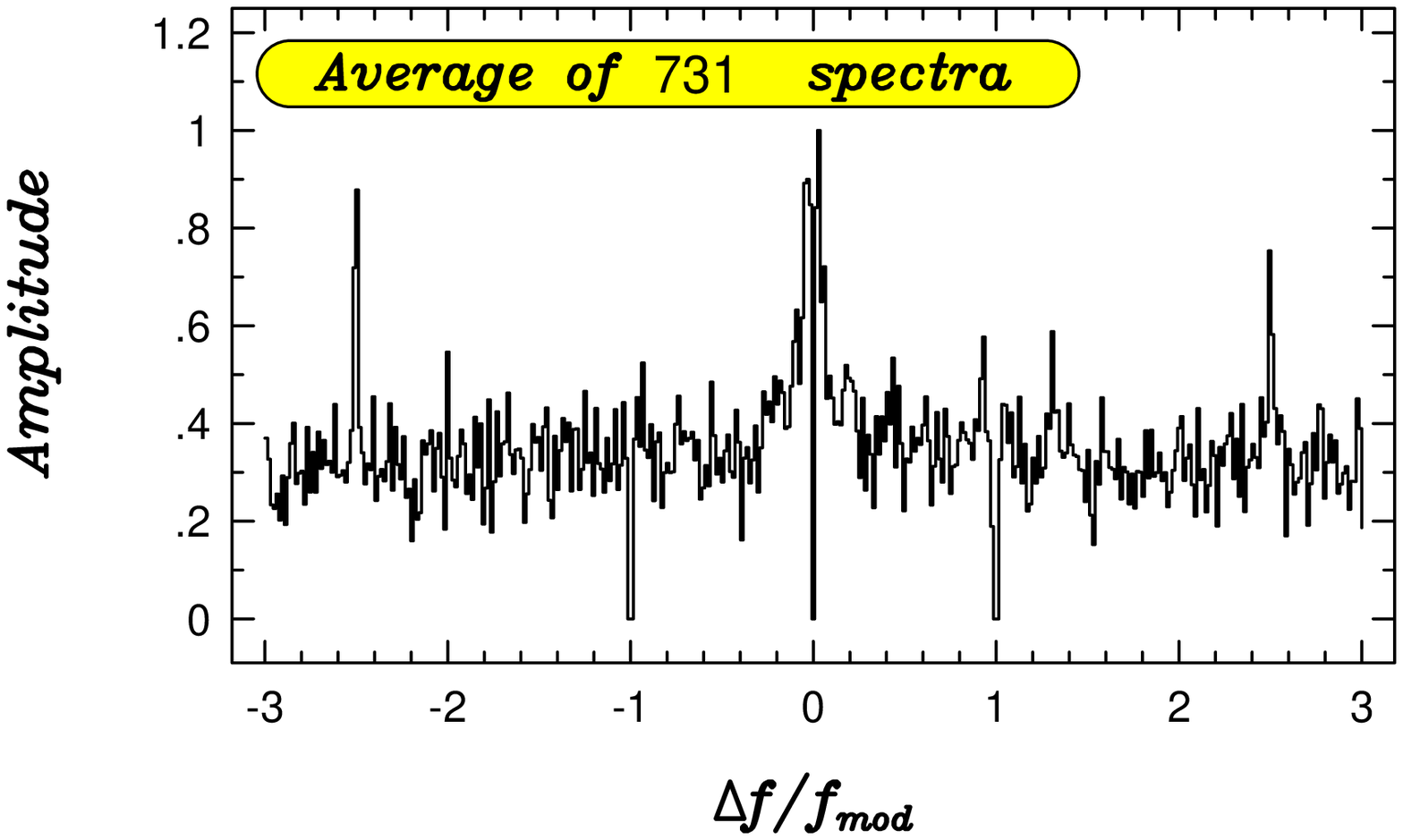,height=50mm,width=90mm}} 
\figcaption[h]{As in Figure~14 but with a {\bf sinusoidal tracer} 
of amplitude of $0.007$~mag added to the signals at 
$\Delta f/f_{mod}=\pm 2.5$.}
\vskip 3mm 

In order to estimate an upper limit for possible quadruplet components, 
we injected tracer signals into the individual time series in the same 
fashion as in our analysis for the $f_m$ component. The effect of an 
$0.007$~mag amplitude tracer is shown in Figure~15. We see that this 
type of component would have been easily detectable from the present 
database. A test with an even lower amplitude of $0.005$~mag shows 
that this would be also detectable, although with a lower significance. 
Therefore, we can safely conclude that if equidistant quadruplet 
structures exist in the BL stars, they would have amplitudes (on average) 
less than $0.004$~mag. 

%
%

\section{CONCLUSIONS}

We have studied the global statistical properties of amplitude- and 
phase-varying fundamental mode RR~Lyrae (BL) stars by performing
a frequency analysis of 6391 variables from the {\sc macho} LMC 
database. We found 731 BL stars which is by far the largest population 
known in a single stellar system. Because of the large sample and the 
long time baseline of the {\sc macho} Project observations, it was 
possible to obtain statistically significant results, describing 
certain overall properties of the BL phenomenon. The main results are 
the following:
\begin{itemize}
\item
Fundamental mode (RR0) BL stars constitute 11.9\% of the presently 
analyzed LMC RR0 population (6158 stars). This incidence rate 
is three times larger than that of their first overtone (RR1) 
counterparts (see Alcock et al. 2000). 
\item
A continuous transition seems to exist between variables showing 
an equidistant triplet around (and including) their main pulsation 
component in their frequency spectra and those displaying only a 
close doublet (BL2- and BL1-types, respectively). This suggests 
that strongly asymmetric modulation amplitudes are not rare among 
the BL2 stars and their BL1 counterparts represent only extreme 
cases when the missing component has such a low amplitude that it 
becomes buried by the observational noise.
\item
74\% of the BL stars have larger modulation amplitudes on the higher
frequency side of the main pulsation component.
\item
Statistical tests indicate that some fraction of the BL2 stars may 
have non-equidistant frequency spacing. There is also a small subgroup 
of 20 variables (RR0$-\nu$M class) which display several close peaks 
at the main pulsation component but without apparent regularity in 
their frequency spacing.
\item
There is a nearly uniform distribution of Blazhko modulation periods 
($P_{BL}$) from $\approx 50$ to several hundred or even thousand 
days. In comparison, variables with $P_{BL}<30$~d are rare.
\item
Modulation amplitude ($A_{+}$ and $A_{-}$) components around the 
main pulsation component ($A_1$) in the frequency spectra are 
usually in the range $0.1 < A_{\pm}/A_1 < 0.3$, but there are 
also more extreme values.
\item
There is a variation with the overall amplitude of $\approx 0.006$~mag 
with a period of $P_{BL}$ in the average brightness of the BL stars.
\item
No clear trace of a quintuplet structure is found around the main 
pulsation component. The overall upper limit of the amplitude of
such a component is estimated to be less than $0.004$~mag. 
\item
No difference is found in the long-term average brightness between 
the singly-periodic and BL RR0 stars. However there are small, 
but systematic, differences in the Fourier decompositions of the 
modulation-free light curves of the BL stars and those of the 
singly-periodic stars. The differences yield average BL light 
curves which have smaller amplitudes and lower degree of skewness 
than their singly-periodic counterparts. 
\item
The PC variables (stars with unresolved power at the main pulsation 
component) show an incidence rate of 2.9\%, which is much lower than 
the rate of 10.6\% of the RR1-PC stars.    
\end{itemize}   

Most of the above conclusions are very similar to those given by 
Moskalik \& Poretti (2003) for the Galactic bulge RR~Lyrae stars. 
The main difference is that they find a much higher incidence rate 
for RR0-BL stars. Their rate is 23\%, which is in the often-quoted 
range for the Blazhko stars (Szeidl 1988), but a factor of 
two higher than what we obtained for our LMC sample. Since the 215 
RR0 stars analyzed by Moskalik \& Poretti (2003) can be considered 
as a statistically significant sample, we tend to agree with them 
that metallicity is the most probable agent for causing the 
difference in the incidence rates. 

Very recently Soszy\'{n}ski et al. (2003) published a similar 
analyisis on 7612 RR~Lyrae stars in the LMC from the {\sc ogle-ii} 
database. Although a detailed comparison of our study with theirs is 
beyond the scope of the present paper, we note the following. Their 
incidence rates for the BL stars are 15\% and 6\% for the RR0 and RR1 
stars, respectively. Both of these rates are somewhat larger than ours. 
It is unclear at this moment what is the reason of this slight discrepancy. 
It may come either from the higher signal-to-noise ratio obtained from 
the use of Differential Image Analysis, or from the higher density of 
variables in the LMC bar of the {\sc ogle} sample, or from the details 
in the search for the modulation components (for example, a search in 
a narrower frequency band around the main pulsation component may lead 
to higher detection rates -- see Sect.~2). Further comparative studies 
are needed to answer these questions.   

Unfortunately, neither the oblique magnetic rotator (Shibahashi 2000) 
nor the rotating resonant pulsator models (Van Hoolst, Dziembowski 
\& Kawaler 1998; Dziembowski \& Cassisi 1999; Nowakowski \& Dziembowski 
2001) are capable of explaining the observed properties of the BL 
population. One major problem for these models is the production of 
strongly asymmetric modulation components. No hint of this behavior 
is found in the oblique magnetic rotator model, whereas the resonant 
one is able to explain only mild asymmetry. Although no current 
observational effort is known to be attempting to detect magnetic 
fields in RR~Lyrae stars, the last result in this field by Chadid 
(2001) does not support the presence of strong magnetic field in 
RR~Lyrae itself. The situation is also complicated concerning the 
direct detection of nonradial pulsation through surface velocity 
field (Kolenberg 2002). 

We conclude that the underlying reason for Blazhko behavior is still
not understood. The large sample reported here will allow investigators
to select subsamples with common features for further study and will
hopefully lead to a theoretical understanding of this rather obvious
observational phenomenon. Precise spectroscopic observations to 
disentangle the large radial pulsation velocity field and the small 
nonradial one, together with a definite measurement of an upper limit 
on the magnetic field, are among the most important observational tasks 
for the near future.   

%
%

\acknowledgements
We are very grateful for the skilled support given our project 
by the technical staff at the Mt. Stromlo Observatory. This work was 
performed under the auspices of the U.S. Department of Energy, National 
Nuclear Security Administration by the University of California, 
Lawrence Livermore National Laboratory under contract No. W-7405-Eng-48, 
the National Science Foundation through the Center for Particle 
Astrophysics of the University of California under cooperative 
agreement AST-8809616, and the Mount Stromlo and Siding Springs 
Observatory by the Bilateral Science and Technology Program of the 
Australian Department of Industry, Technology and Regional Development. 
KG acknowledges a DOE OJI grant, and CWS and KG thank the Sloan 
Foundation for their support. DLW was a Natural Sciences and 
Engineering Research Council (NSERC) University Research Fellow 
during this work. DM is supported by FONDAP Center for Astrophysics 
15010003. The support of {\sc otka t$-$038437} is acknowledged.

%
%

\end{document}